\documentclass{jfm}

\usepackage{graphicx}
\usepackage{newtxtext}
\usepackage{newtxmath}
\usepackage{natbib}
\usepackage{hyperref}
\hypersetup{
    colorlinks = true,
    urlcolor   = blue,
    citecolor  = black,
}

\newcommand{\RomanNumeralCaps}[1]
\linenumbers

\usepackage[dvipsnames]{xcolor}
\usepackage{amsmath}
\usepackage{amsfonts}
\usepackage{mathrsfs}
\usepackage{amssymb}
\usepackage{relsize}
\usepackage{cancel}
\usepackage{siunitx}
\usepackage{float}
\usepackage{enumitem}
\usepackage{tikz}
\usepackage{pgfplots}
\pgfplotsset{compat=1.18}
\usepackage{import}
\usepackage{verbatim}
\usepackage{ulem}

\renewcommand{\eqref}[1]{(\ref{#1})}
\renewcommand{\vec}{\boldsymbol}

\newcommand{\diver}[1]{\vec{\nabla} \cdot #1}
\newcommand{\grad}[1]{\vec{\nabla} #1}
\newcommand{\lapl}[1]{\nabla^2 #1}

\newcommand{\abs}[1]{\lvert #1 \rvert}%
\newcommand{\norm}[1]{\left \lVert #1 \right \rVert}%
\newcommand{\derpar}[2]{\frac{\partial #1}{\partial #2}}

\newcommand{\derparm}[3]{\frac{\partial^2 #1}{\partial #2 \partial #3}}

\newcommand{\parton}[1]{\left ( #1 \right )}
\newcommand{\parq}[1]{\left [ #1 \right ]}
\newcommand{\parket}[1]{\left \langle #1 \right \rangle}

\newcommand{\Ubar}{\overline{U}}
\newcommand{\Prod}{\mathcal{P}}
\newcommand{\Diss}{\mathcal{D}}
\newcommand{\bigint}{\mathlarger{\mathlarger{\int}}}

\title{Large-scale streak instabilities of transitional channel flow}

\author{N. Ciola\aff{1,3}
  \corresp{\email{n.ciola@phd.poliba.it}},
  Y. Duguet\aff{2},
  J.-C. Robinet\aff{1},
  P. De Palma\aff{3},
 \and S. Cherubini\aff{3}}

\affiliation{\aff{1}DynFluid, Arts et Métiers Paris /CNAM, 151 Bd de l’Hôpital, 75013 Paris, France
\aff{2}LISN-CNRS, Université Paris-Saclay, 507 Rue du Belvédère, 91405 Orsay, France
\aff{3}DMMM, Politecnico di Bari, Via Re David 200, 70125 Bari, Italy}

\begin{document}
\maketitle

\begin{abstract}

The emergence of large-scale spatial modulations of turbulent channel flow, as the Reynolds number is decreased,  is addressed numerically using the framework of linear stability analysis. Such modulations are known as the precursors of laminar-turbulent patterns found near the onset of relaminarisation. A synthetic two-dimensional base flow is constructed by adding finite-amplitude streaks to the turbulent mean flow. The streak mode is chosen as the leading resolvent mode from linear response theory. Besides, turbulent fluctuations can be taken into account or not by using a simple Cess eddy viscosity model. The linear stability of the base flow is considered by searching for unstable eigenmodes with wavelengths larger than the base flow streaks. As the streak amplitude is increased in the presence of the turbulent closure, the base flow loses its stability to a large-scale modulation below a critical Reynolds number value. The structure of the corresponding eigenmode, its critical Reynolds number, its critical angle and wavelengths are all fully consistent with the onset of turbulent modulations from the literature. The existence of a threshold value of the Reynolds number is directly related to the presence of an eddy viscosity, and is justified using an energy budget. The values of the critical streak amplitudes are discussed in relation with those relevant to turbulent flows.
\end{abstract}

\maketitle


\section{Introduction}

Any description of a turbulent flow starts with the identification
of the most relevant coherent structures and of their dynamics. 
For shear flows in the vicinity of solid walls,
streamwise streaks, i.e. spanwise modulations of the streamwise velocity, together with the associated streamwise vortices,
are by far the most reported coherent structures \citep{kline1967structure}. Streamwise streaks do not emerge
from the laminar flow as an instability but rather as a manifestation of the non-normal amplification of perturbations induced by streamwise vortices that could be already present in the associated laminar shear flow \citep{schmid2002stability}. The distribution of the spanwise widths of streaks peaks at a wavelength close to 100 wall units \citep{kline1967structure}. This scale is associated with the immediate vicinity of the wall, however in this study we are mostly concerned with marginally low Reynolds number values for which streaks and vortices extend across the channel gap, making the distinction between near-wall and outer region insignificant. Streamwise streaks are themselves linearly unstable to various types of velocity perturbations with different possible lengthscales \citep{schoppa2002coherent} and spatial symmetries \citep{andersson2001breakdown}, see \citet{lozano2021cause} for a review of such instabilities in the literature. At the scale of the streaks themselves, streak instabilities generate additional vorticity that feedbacks on the streamwise rolls and prevents them from viscously decaying \citep{hamilton1995regeneration,waleffe1997self}. 
This nonlinear mechanism, known as the self-sustaining process (SSP), is local in nature and explains the existence of non-laminar flow, including turbulence, finite-amplitude unsteady perturbations, and edge states \citep{waleffe1997self}.
Streak instabilities at larger scales have also been reported  as one possible explanation for the occurence of large-scale motions (LSMs) frequently reported in high-Reynolds number experiments and simulations \citep{jimenez2013near, de2017streak}. Such an instability scenario coexists with a transient growth mechanism justifiying LSMs as the flow disturbances most amplified by non-normal effects \citep{del2006linear,pujals2009note}. Interestingly, high-Reynolds-number flows have always been a topic of investigation in the shear flow literature, as justified by numerous real world applications. However, at the lowest Reynolds numbers where turbulence exists,  the structure of the turbulent flow features a different class of large-scale coherent structures, which is so far poorly understood. Focusing on the example of channel flow between two infinitely large parallel plates and driven by a pressure gradient, as the friction Reynolds number ($Re_{\tau}$) is decreased just below $95 \pm 1$, the turbulent flow begins to display modulations with an oblique wavevector \citep{shimizu2019bifurcations,kashyap2020flow}, interpreted recently as a linear  instability of the turbulent regime \citep{kashyap2022linear}. 
For lower $Re_{\tau}$, the amplitude of the modulations increases and the weaker zones relaminarise, leading to laminar-turbulent patterns \citep{tsukahara2005dns,tuckerman2014turbulent,tuckerman2020patterns}. For even lower $Re_{\tau}$, the patterns break into solitary bands of turbulence \citep{shimizu2019bifurcations}. A similar phenomenology was reported in most wall-bounded flows, yet sometimes without a clear evidence for the modulational stage as {\it e.g.} in plane Couette flow \citep{gome2023patterns}. With the hope that the simplicity of the spatial modulations is likely to hide a simpler phenomenology, we choose to investigate the channel case.
We focus hence in this article on the modulational regime of turbulent channel flow and aim at predicting quantitatively such modulations using direct numerical simulation and tools from linear stability analysis (LSA). \\

Applying hydrodynamic stability concepts to a turbulent base flow poses serious conceptual challenges. In principle, in the deterministic context
of the Navier--Stokes equations, linearisation is carried out in the neighbourhood of some relevant equilibrium flow state characterised by a temporal symmetry, {\it e.g.} a steady state, a travelling wave or a relative periodic orbit. Such flow states should be exact solutions of the governing equations. 
In shear flows, such self-sustained states always feature three-dimensional streamwise streaks that are linearly unstable \citep{waleffe2001exact,kawahara2012significance}. 
A few recent attempts to link the occurence of laminar-turbulent patterning to bifurcations of such three-dimensional solutions of the Navier--Stokes equations were reported in the context of pipe flow \citep{chantry2014genesis} and plane Couette flow \citep{reetz2019exact}. In the channel flow geometry, several solutions akin to patterning were also reported recently \citep{paranjape2020oblique,paranjape2023direct} but they were so far not linked to spatially unmodulated solutions. Given that the exact dynamical relevance of such solutions to the turbulent dynamics remains subtle \citep{cvitanovic2013recurrent}, we find it here preferable to focus on a different approach that actively takes into account the presence of turbulent fluctuations. Linearising around a chaotic attractor using concepts such as Lyapunov exponents and covariant Lyapunov vectors \citep{inubushi2015regeneration} is such a possibility, albeit a computationally heavy one. Taking into account the very fine details of the base flow, both spatially and temporally, results anyway in an increasingly complex analysis from which no simple physical interpretation might emerge. A recent attempt
to extract the dominant large-scale instability using this approach in plane Couette flow did not lead to a prediction of modulations characterised by an oblique wavevector \citep{ishikawa2018orbital}.  Instead, we adopt here a different theoretical approach, based on the choice for a simpler base flow, where statistical symmetries are as much as possible taken into account.\\

In the last decade, hydrodynamic stability calculations have began to become relevant for the generation of coherent structures within a turbulent context, once it was realised that LSA could be carried out, in practice, in the vicinity of the turbulent {\it mean flow} rather than the laminar base flow. Because of the invariance of the problem with respect to translations in time and in the two planar directions, the mean flow, understood in practice as the time-averaged velocity field, depends only on the wall-normal coordinate $y$, but not on the streamwise coordinate $x$, the spanwise coordinate $z$, or the time $t$. Intriguingly, this approach fails systematically in channel flow : the mean flow of turbulent channel flow was reported to be linear stable to all perturbations, both for high \citep{reynolds1967stability} and low values of $Re_{\tau}$ \citep{kashyap2024linear}. A glance at any velocity field in the turbulent regime  (see e.g. Fig. 1 in  \cite{kashyap2022linear}) reveals the ubiquity of long-lasting streamwise streaks. The relatively rapid emergence time of the large-scale modulations, comparable to the viscous timescale, suggests that the y-dependent mean flow is poorly relevant as a base flow
for LSA in the present context. Indeed, from most visualisations it is clear that the large-scale modulations grow on a streak field rather than on a bare, structure-free base flow. We hence revisit the linearisation of \cite{kashyap2025laminar} by adding well-controlled streamwise streaks to the laminar base flow,
and by allowing for linear instabilities with lengthscales larger than the  characteristic size of streaks. One of the motivations for including streamwise streaks is their coherence, {\it i.e.} their presence in the mean flow if the time average holds over a finite time only. Finite-time averaging is however no rigorous proper Reynolds averaging, and linear LSA around such a base flow is ambiguous to interpret theoretically.\\
Because of the turbulent context, fluctuations and their nonlinear self-interaction can also not be ignored. Reynolds stresses must be accounted for and modelled. The quest for an optimal parametrisation of turbulent closures in channel flow is outside the scope of this study. We focus here on a simple and popular eddy viscosity closure, already used in former mean flow stability computations \citep{del2006linear,pujals2009note,morra2019relevance}, which we can decide to either include or not in the analysis. Although this eddy viscosity model is simplistic, as we shall see it already leads to meaningful conclusions regarding the possibility for large-scale instabilities as precursors of turbulent modulations and laminar-turbulent patterning.\\

This paper is structured as follows. Section 2 contains the flow governing equations and the primary resolvent analysis from which synthetic streaks are computed. The results of the LSA of the streaky base flow are shown in Section 3. Section 4 contains an energy budget analysis applied to the unstable mode found. The model assumptions are discussed in light of the results in Section 5. Conclusions and outlooks are eventually given in Section 6. \\ 

\begin{figure}
\centering
\raisebox{2.3in}{(a)} \includegraphics[width=0.9\textwidth]{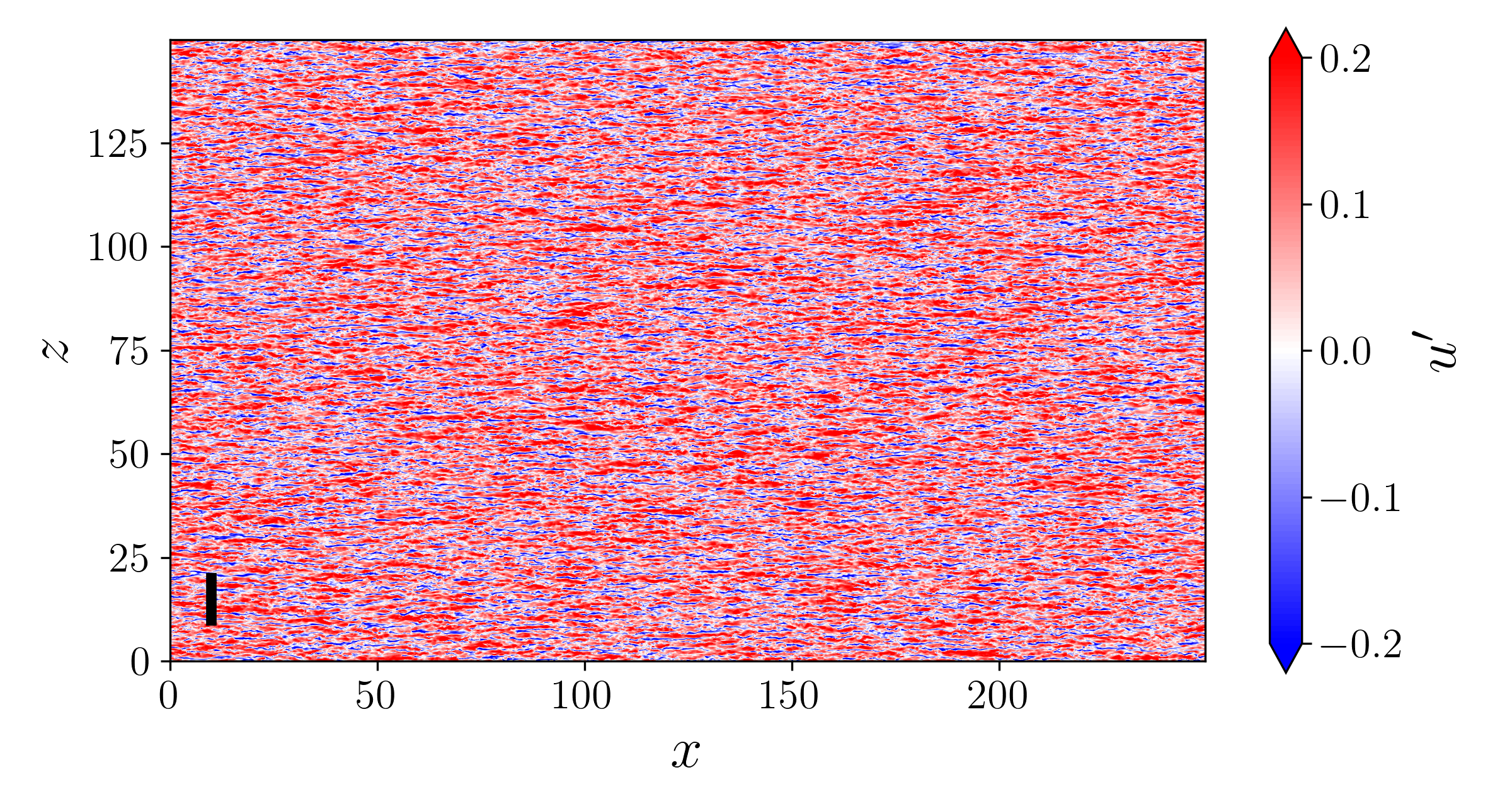}\\
\raisebox{2.3in}{(b)} \includegraphics[width=0.9\textwidth]{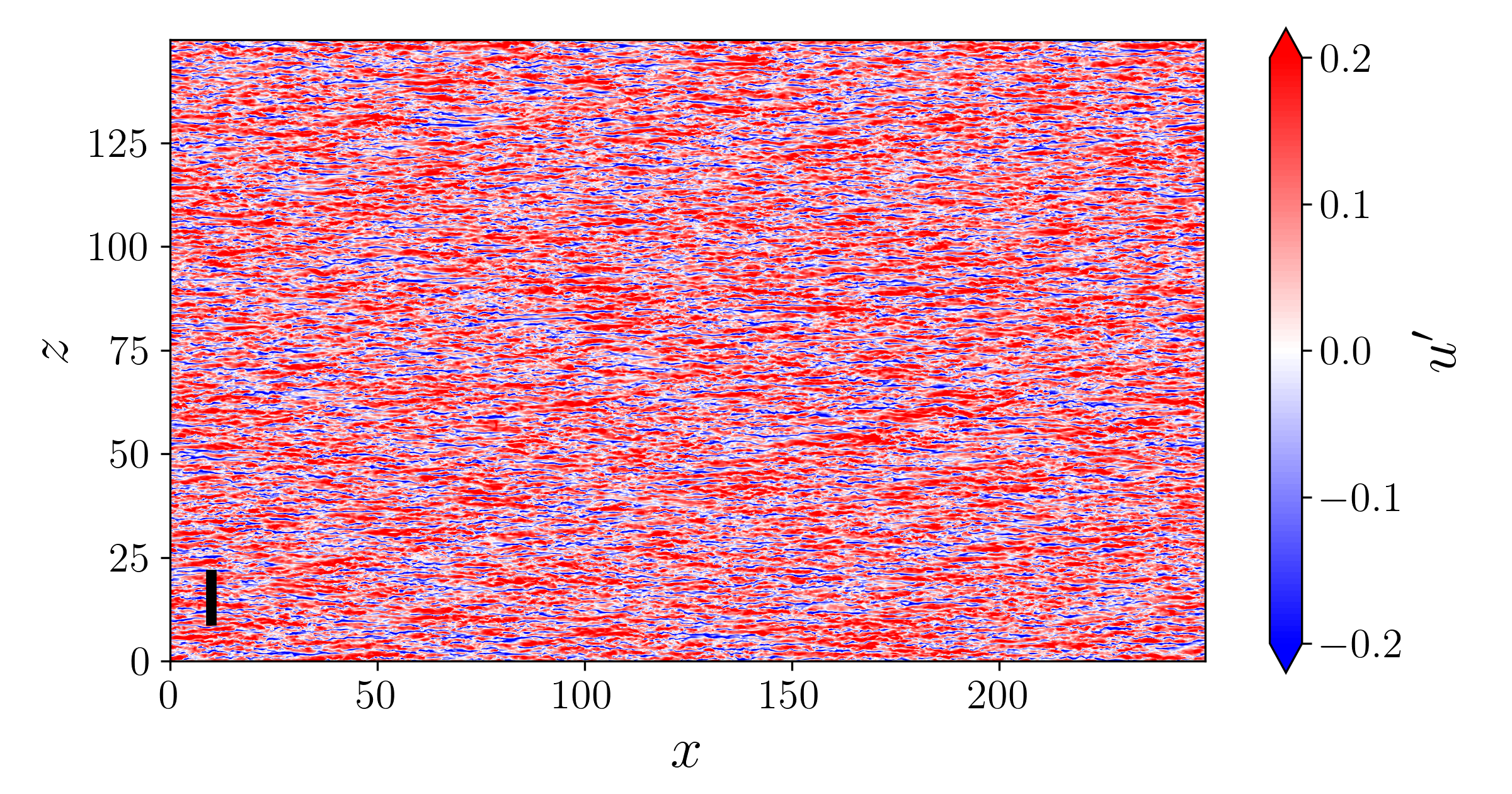}\\
\raisebox{2.3in}{(c)} \includegraphics[width=0.9\textwidth]{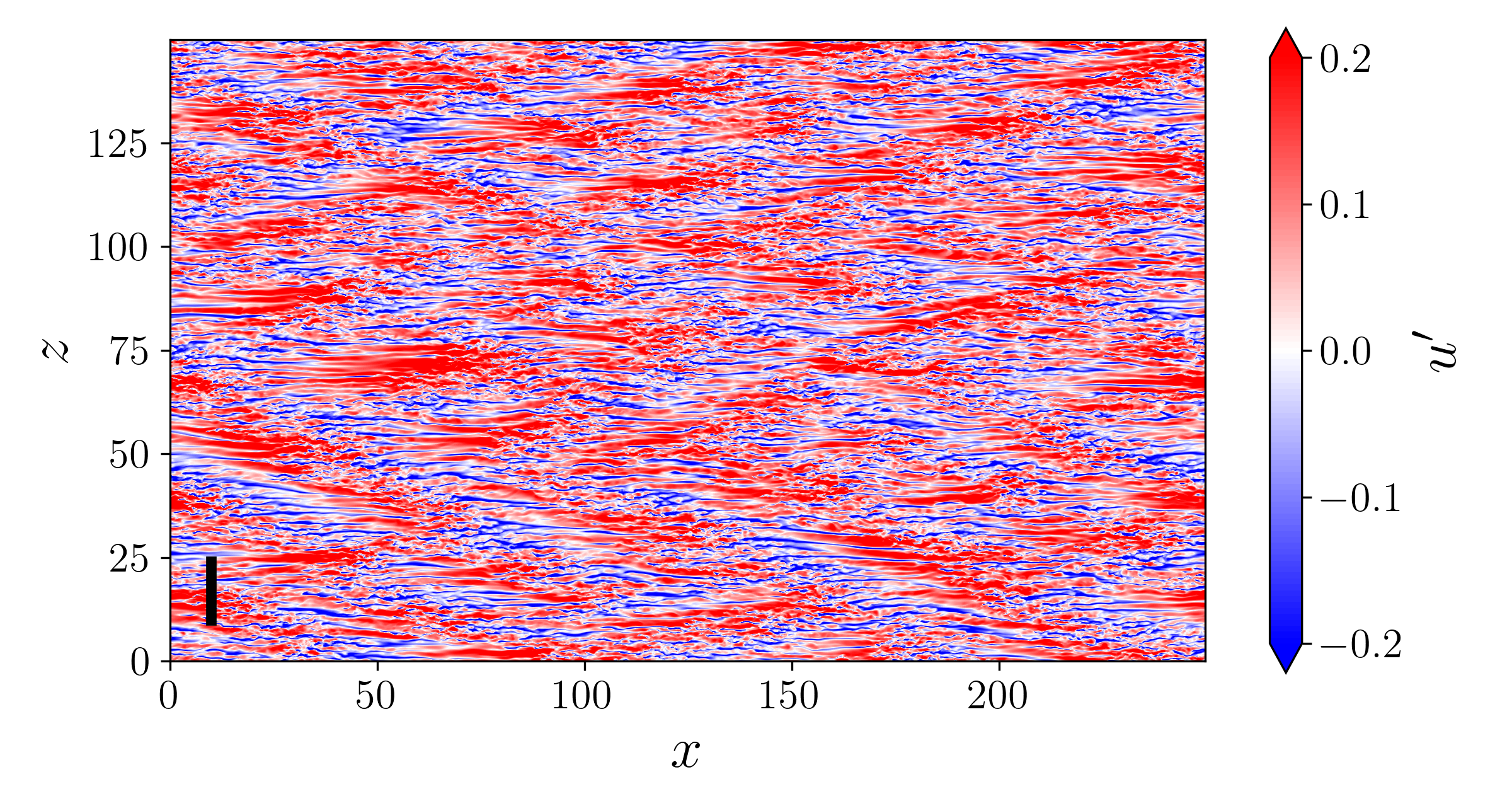}
\caption{Contours of streamwise velocity fluctuations in the plane $y^+\approx 35$ from a DNS at (a) $Re_{\tau} = 98$, (b) $Re_{\tau} = 92$ and (b) $Re_{\tau} = 71$  ($y\approx0.36$ for $Re_{\tau}=98,92$ and $y\approx0.49$ for $Re_{\tau}=71$). The black vertical segment at the bottom left corner of each panel indicates a spanwise length of $1000$ wall units.
}
\label{fig:dns_slice_xz}
\end{figure}


\section{Flow setup and resolvent modes}
\label{sec:resolvent}

\subsection{Governing equations}

We consider the incompressible flow in a channel between two infinitely extended parallel no-slip walls. The flow is driven by a time-varying forcing (streamwise pressure gradient) such that the streamwise flow rate is held constant. The spanwise flow rate is also imposed constant to be zero. The streamwise, wall-normal and spanwise directions are denoted, respectively, by $x$, $y$ and $z$. Velocity components in these directions are denoted, respectively, by $u$, $v$ and $w$. Quantities without any superscript have been non-dimensionalised by the channel half-gap $h$ (such that  $0\le y \le 2$) and the (imposed) channel bulk velocity $U_b$ ($U_b=\int_0^2\int_0^{L_z} u \; dydz / 2L_z$). Quantities with a $+$ superscript have been non-dimensionalised by the viscous lengthscale $\delta_{v} = \nu/u_{\tau}$ and the friction velocity $u_{\tau}=\sqrt{\tau_w/\rho}$, where $\rho$ is the fluid density, $\nu$ the kinematic viscosity and $\tau_w$ the (measured) mean wall shear stress.\\

The instantaneous flow is governed by the  Navier-Stokes equations for incompressible flows
\begin{equation}
\label{eq:navierstokes}
\begin{cases}
\displaystyle \derpar{\vec{u}}{t} + \vec{u}\cdot\grad{\vec{u}} = - \grad{p} + \frac{1}{Re} \lapl{\vec{u}} + \vec{f}_b, \\[1ex]
\displaystyle \diver{\vec{u}}=0,
\end{cases}
\end{equation}
where $Re=U_bh/\nu$ is the Reynolds number and $\vec{f}_b$ is the time-varying forcing that keeps the flow rate constant.\\

Direct Numerical Simulation (DNS) of equations \eqref{eq:navierstokes} performed in this study makes use of the \textsf{channelflow 2.0} code \citep{gibson2021}. It is based on Fourier decomposition in the streamwise ($x$) and spanwise ($z$) directions and Chebyshev collocation in the wall-normal direction ($y$). Dealiasing in the wall-parallel directions is performed using the $2/3$ rule. A third-order-accurate backward finite-difference scheme is used for time discretisation. The simulations illustrated in figure \ref{fig:dns_slice_xz} are performed in a domain of size $L_x \times L_y \times L_z = 250\times2\times150$,  in the streamwise, wall-normal and spanwise directions, respectively, in units of the channel half-gap. The collocation points (before dealiasing) are $1500\times65\times1800$, respectively leading to resolutions comparable to previous works \citep{kashyap2020flow,kashyap2022linear,shimizu2019bifurcations}. Equivalent resolutions are used for the other domains mentioned in the paper. The control parameter in our simulations is the Reynolds number based on the bulk velocity. However, results are presented as a function of the friction Reynolds number $Re_{\tau}=u_{\tau}h/\nu$, which can be computed \textit{a posteriori}, to ease comparison with previous works.\\

At the lowest Reynolds number considered here ($Re_{\tau}=71$), the turbulent flow appears as a pattern of oblique bands (figure \ref{fig:dns_slice_xz} (c)). The alternance of extended blue and red regions indicates the presence of the large-scale flow which coexists with the pattern \citep{duguet2013oblique}. At a higher value of $Re_{\tau}=92$, such a pattern of alternatively laminar and turbulent regions is replaced in DNS by a weaker modulation of a more standard streaky turbulent flow (figure \ref{fig:dns_slice_xz} (b)). Instead, for $Re_{\tau}$ above 95, turbulence appears in DNS unmodulated by large-scale oblique structures  (figure \ref{fig:dns_slice_xz} (a)), consistently with \cite{kashyap2022linear},  and will be qualified as {\it featureless} after \citet{shimizu2019bifurcations}.

\subsection{Streaks computation by resolvent analysis}

As mentioned in the introduction, DNS visualisations suggest that the modulation of the turbulent channel flow, reported by \citet{kashyap2022linear}, exists on top of a field of predominantly streamwise streaks, see figure \ref{fig:dns_slice_xz}. The objective of this work is to investigate this phenomenology by performing a modal stability analysis of streaks superimposed on the mean shear flow. A wealth of previous studies have demonstrated that typical near-wall structures can be computed using the Navier-Stokes operator linearised around the mean flow either by considering the harmonic forcing problem \citep{hwang2010linear,mckeon2010,sharma2013coherent,moarref2013} or the initial value problem \citep{del2006linear,pujals2009note}. The results of this analysis are briefly revised here at low $Re$.

Using the Reynolds decomposition, the solution of \eqref{eq:navierstokes} can be partitioned as the sum of the mean flow $\vec{\Ubar} = \Ubar(y)\vec{e}_x$ plus fluctuations. Then, following \citet{reynolds1972}, the turbulent fluctuations are further partitioned into a coherent part and an incoherent part. Neglecting the interactions between the two parts, the nonlinear interaction of the coherent part with itself and modelling the incoherent fluctuations using a classical eddy viscosity model, one obtains the following linear system
\begin{equation}
\label{eq:eddy-resolvent}
\begin{cases}
\displaystyle \derpar{\vec{u}^{\prime}}{t} + \vec{\Ubar}\cdot\grad{\vec{u}^{\prime}} + \vec{u}^{\prime}\cdot\grad{\vec{\Ubar}} + \grad{p^{\prime}} - \frac{1}{Re} \lapl{\vec{u}^{\prime}} - \diver{\parq{\nu_t\parton{\grad{\vec{u}^{\prime}} + (\grad{\vec{u}^{\prime}})^T}}} = \vec{f}, \\[1ex]
\displaystyle \diver{\vec{u}^{\prime}}=0,
\end{cases}
\end{equation}
for the coherent velocity fluctuation $\vec{u}^{\prime}$ (resp. $p'$ for the coherent pressure fluctuation), where $\vec{f}$ is a generic forcing term (\textit{cf.} \citet{morra2019relevance}). The coherent fluctuation is interpreted mainly as the large scales, but it also includes the streaks because of their temporal and spatial coherence. Smaller scales as well as the remaining other flow structures, that by default we consider as "incoherent", are not explicitely resolved, instead they are modelled as an eddy viscosity. 

In this work, as in previous works on the channel flow \citep{del2006linear,pujals2009note,hwang2010linear,park2011,alizard2015linear,ciola2024large}, the \citet{cess1958} eddy viscosity formula, as reported by \citet{reynolds1967stability}, is employed:
\begin{equation}
\label{eq:cess_formula}
    \nu_t(y) = \frac{1}{2Re}\left \{ \parq{1 + \parton{\frac{Re_{\tau}k}{3} \parton{2y-y^2} \parton{3-4y+2y^2} \parton{1-e^{\parton{\abs{y-1}-1}Re_{\tau}/A}} }^2 }^{1/2} - 1 \right \},
\end{equation}
with $k=0.426$, $A=25.4$ \citep{del2006linear} and $y \in [0,2]$. Recent comparative studies have shown that including an eddy viscosity term in the linear operator is crucial for quantitatively improved predictions of the linear model \citep{morra2019relevance,illingworth2018,symon2023use}. In this study, we will compare stability results with and without eddy viscosity. Borrowing the terminology from \citet{morra2019relevance}, equation \eqref{eq:eddy-resolvent} without the eddy viscosity term will be referred as the quasi-laminar model, whereas the full equation will be referred as eddy viscosity model. Note, however, that in both cases the eddy viscosity given by equation \eqref{eq:cess_formula} is used to compute the mean profile solving the one-dimensional boundary value problem of \citet{reynolds1967stability}.\\

\begin{figure}
\centering
\raisebox{1.3in}{(a)}\includegraphics[width=0.21\textwidth]{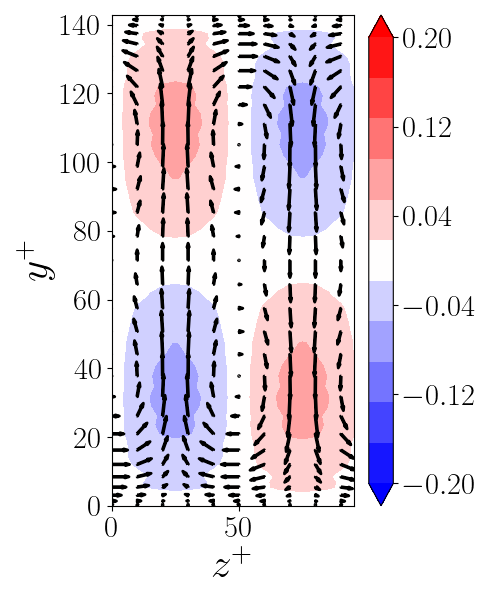}
\raisebox{1.3in}{(b)}\includegraphics[width=0.21\textwidth]{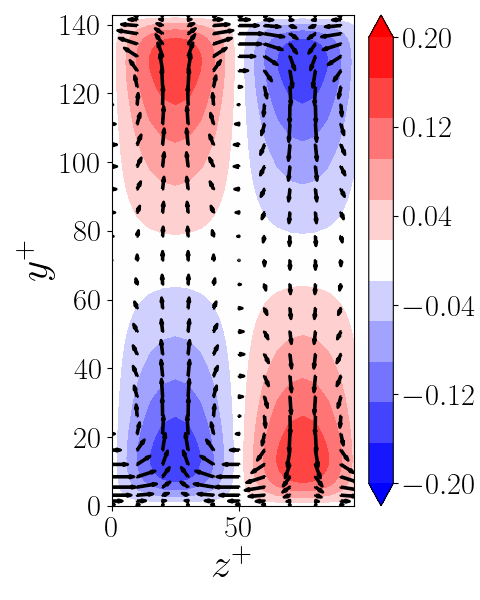}
\raisebox{1.3in}{(c)}\includegraphics[width=0.21\textwidth]{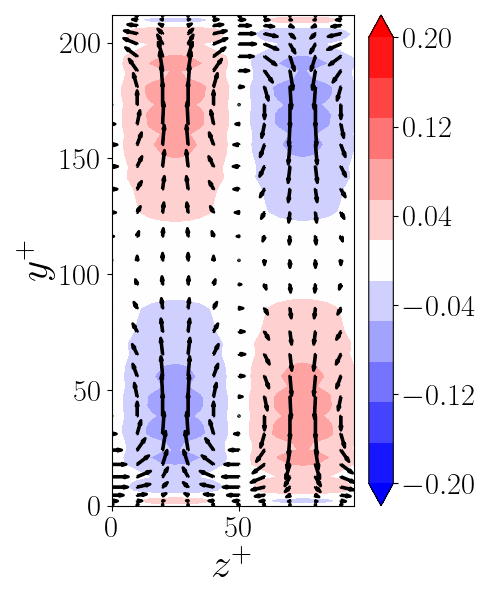}
\raisebox{1.3in}{(d)}\includegraphics[width=0.21\textwidth]{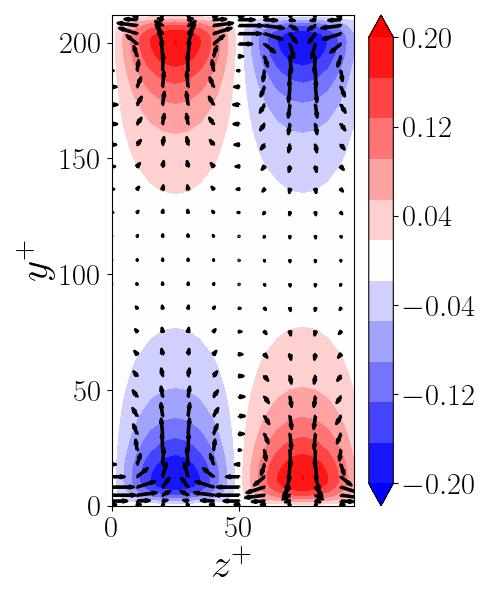} \\
\raisebox{1.3in}{(e)}\includegraphics[width=0.21\textwidth]{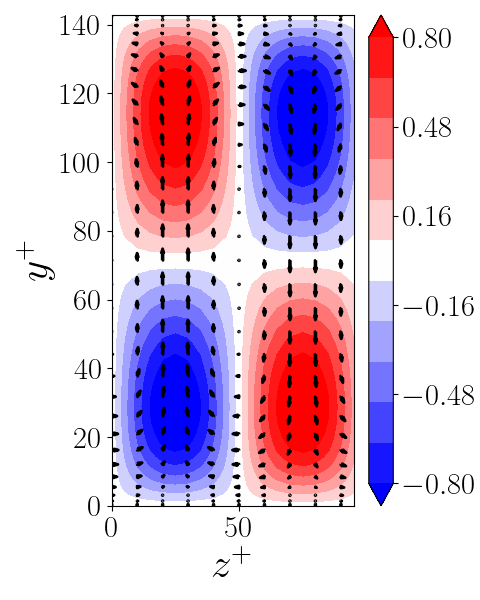}
\raisebox{1.3in}{(f)}\includegraphics[width=0.21\textwidth]{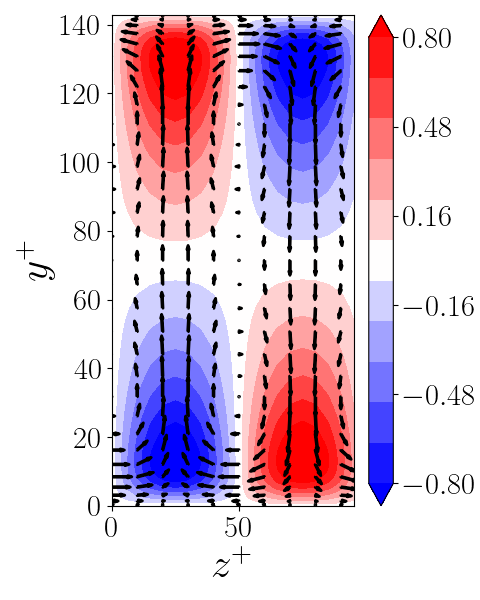}
\raisebox{1.3in}{(g)}\includegraphics[width=0.21\textwidth]{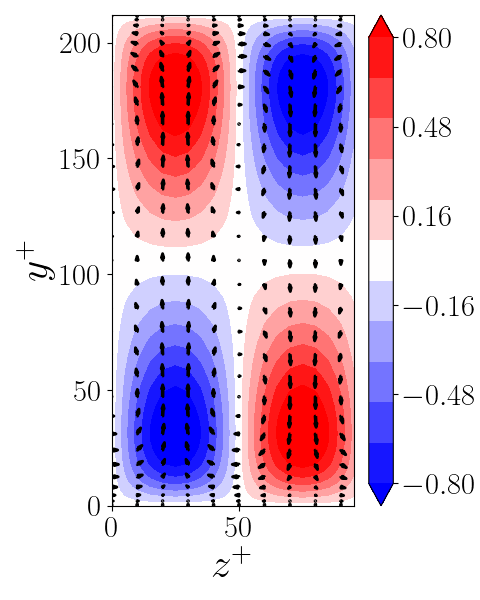}
\raisebox{1.3in}{(h)}\includegraphics[width=0.21\textwidth]{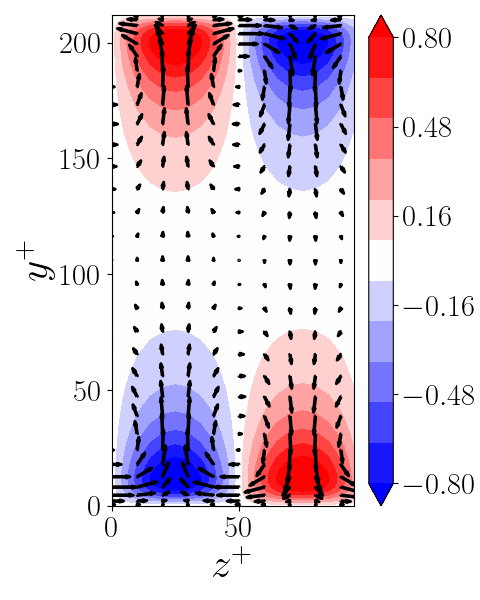}
\caption{Optimal harmonic forcing (a-b-c-d) and response velocity field (e-f-g-h) obtained from the resolvent analysis of the mean flow for (a-b-e-f) $Re_{\tau}=71$ and (c-d-g-h) $Re_{\tau}=105$. In all panels contours denote the streamwise component while quivers stand for transverse components. The scale of the arrows in the top row is ten times larger than in the bottom row. Streamwise uniform case ($k_x=0$). (a-c-e-g) Quasi-laminar model; (b-d-f-h) eddy viscosity model.
}
\label{fig:resolvent_modes}
\end{figure}

The linear nature of the system \eqref{eq:eddy-resolvent} greatly simplifies its analysis: by considering the Fourier transform of the equation in the wall-parallel directions and in time, the harmonic response is computed as $\vec{\hat{u}}(y;k_x,k_z,\omega)=\mathcal{H}(y;k_x,k_z,\omega) \vec{\hat{f}}(y;k_x,k_z,\omega)$, where $\mathcal{H}(y;k_x,k_z,\omega)$ is the resolvent operator, the hats denote Fourier transforms in $x$, $z$ and $t$, $\omega$ is the angular frequency of both the forcing and the response, and $k_x$ and $k_z$ are the streamwise and spanwise wavenumbers. For a given $\{\omega,k_x,k_z\}$ triplet. The optimal harmonic forcing is defined as the function $\vec{\hat{f}}(y)$ that maximises the amplitude ratio between response and forcing (cf. \citet{hwang2010linear}):
\begin{equation}
\label{eq:resnorm}
    R(k_x,k_z,\omega) = \max_{\vec{\hat{f}}\neq\vec{0}} \frac{\norm{\vec{\hat{u}}}^2} {\norm{\vec{\hat{f}}}^2}=\norm{\mathcal{H}(y;k_x,k_z,\omega)}^2 
\end{equation}
where the standard $L^2$ norm in $y$ is used both for $\vec{\hat{u}}$ and $\vec{\hat{f}}$. In this study, the optimal harmonic forcing problem is solved using an in-house code which performs the Singular Value Decomposition of the discretised resolvent operator \citep{schmid2014,symon2018}. After Fourier transform in $x$, $z$ and $t$ and discretisation in $y$, the system \eqref{eq:eddy-resolvent} can be written as $\parton{\iota \omega \mathsfbi{B} - \mathsfbi{L}}\mathsfbi{\hat{q}}=\mathsfbi{\hat{f}}$, where $\mathsfbi{\hat{q}}$ contains the four primitive variables $\hat{u}$, $\hat{v}$, $\hat{w}$ and $\hat{p}$ for each point of the computational grid, $\mathsfbi{B}$ is the prolongation matrix \citep{cerqueira2014eigenvalue}, $\mathsfbi{L}$ is the discretised linear Navier-Stokes operator and $\iota$ the imaginary unit. The discretised resolvent operator can be computed by a matrix inversion as $\mathsfbi{H}=\parton{\iota \omega \mathsfbi{B} - \mathsfbi{L}}^{-1}$ and decomposed such that $\mathsfbi{H}=\mathsfbi{U}\boldsymbol{\mathsf{\Sigma}}\mathsfbi{V}^T$. The most amplified harmonic forcing is given by the first column of $\mathsfbi{V}$, the corresponding response mode by the first column of $\mathsfbi{U}$ and the energy amplification factor $R$ by the square of the leading singular value of $\boldsymbol{\mathsf{H}}$, which is the first diagonal element of $\boldsymbol{\mathsf{\Sigma}}$. For the discretisation of differential operators, $N_y=65$ Chebyshev collocation points have proven sufficient at the low values of $Re$ considered. The code was validated by reproducing the results of \citet{hwang2010linear}.

Resolvent analysis is not the main focus of this work, it is used here as a means of defining the streaks that are relevant for the stability analysis. Note that there are other alternative ways to define two-dimensional streak modes, including eigenmodes from the associated Stokes operator \citep{waleffe1997self} or data-driven techniques where streaks are directly extracted from DNS \citep{hack2014streak} or experimental data \citep{liu2024}. The present choice of the resolvent modes was motivated by the available machinery together with the need to make the results reproducible. Therefore, at this stage, we do not explore systematically the optimal harmonic response. It is well known that typical near-wall structures in the considered flow are streaks characterised by a dominant spanwise wavelength $\lambda^+_{z,s} \approx 100$, defined in wall units. Moreover, in a first approximation, they can be seen as streamwise-independent structures (\textit{straight streaks} following \citet{liu2024}). Streamwise streaks are routinely explained by the advection of streamwise vortices by the lift-up effect \citep{ellingsen1975,landahl1980}, known as a linear mechanism, although explaining their persistence in turbulent flows requires nonlinear concepts such as the SSP. As it happens, resolvent analysis with the operator linearised around the mean flow is well-suited to compute such structures including the one optimally amplified by linear mechanisms. In the resolvent framework, the forcing $\vec{f}$ represents the vortices and the response $\vec{u}^{\prime}$ the streak field. The optimal harmonic forcing problem is solved here for $k_x = 0$ and $k^+_z = 2\pi/100$ and using $20$ equi-spaced $\omega$ values in $[0,2\pi]$. Optimal amplification is found for steady forcing ($\omega=0$) as in \citet{hwang2010linear}. The problem is solved for a range of $Re$ relevant to the modulational regime of turbulent patterns in the channel, in particular $Re_{\tau}=[71, 78, 84, 91, 95, 98, 105]$. \\

As anticipated, the forcing is found to be mainly directed in the wall-normal and spanwise directions, while the response is primarily dominated by the streamwise velocity component. Figure \ref{fig:resolvent_modes} shows the streamwise component of the forcing and response as shaded contours and the transverse components as arrows in the $y-z$ plane for two values of $Re_{\tau}$. The typical picture of streamwise rolls and induced streaks is recognised here. Comparing the quasi-laminar model (panels a-c-e-g) with the eddy viscosity model (panels b-d-f-h), it is found that the eddy viscosity squeezes both streaks and vortices towards the wall. This effect increases with the Reynolds number. This is likely due to the wall-normal dependence of the eddy viscosity \eqref{eq:cess_formula}. We now turn towards the linear stability analysis of the streaks.


\section{Stability analysis}
\label{sec:lsa}

\subsection{Formulation}

The lift-up effect is only one of the several mechanisms involved in the self-sustaining process of wall turbulence \citep{hamilton1995regeneration,waleffe1997self}. Another step in this cyclic process is the so-called secondary instability of the streaks that have been amplified by the lift-up effect \citep{schoppa2002coherent,park2011,alizard2015linear}. In the classical SSP picture, that stage precedes the feedback onto the streamwise vortices. More recently, \citet{de2017streak} and \citet{ciola2024large} have shown that streak instabilities can also contribute to the generation of larger-scale structures, thereby allowing for an escape from the one-scale-only process described by the original SSP theory. At lower $Re$ typical of the transitional regime, a similar analysis is of interest to investigate whether the secondary instability of streaks can be related to the emergence of large-scale modulations of the turbulent flow suggested from DNS observations.\\

As illustrated in the previous section, the resolvent analysis identifies an optimal response with zero frequency. The velocity field associated with this response is hereafter denoted by $\vec{u}_s$. Due to the linearity of equation \eqref{eq:eddy-resolvent}, $\vec{u}_s$ is defined up to a multiplicative constant. Following our previous work \citep{ciola2024large}, $\vec{u}_s$ is rescaled such that:
\begin{equation}
\label{eq:streaks_amp}
    \frac{\max_{y,z} u_s - \min_{y,z} u_s}{2\Ubar_{c}} = 1,
\end{equation}
where $u_s$ (as a scalar field) is the streamwise velocity component of the streaks field and $\Ubar_c$ is the centreline velocity of the turbulent mean profile. \\

A new base flow $\vec{U}$ is constructed from the knowledge of the mean flow $\vec{\Ubar}$ and the streak field $\vec{u}_s$ 
 as \begin{equation}
    \vec{U} = \vec{\Ubar} + A_s \vec{u}_s,
\end{equation}
with $A_s>0$ interpreted as the \textit{streak amplitude}. Consistently with the previous section, $\vec{U}=(U(y,z),V(y,z),W(y,z))$ is uniform in the streamwise direction, which simplifies  and reduces the computational cost of LSA compared to the case of a genuinely three-dimensional base flow.\\

Since both the mean flow and the optimal response are steady, we expect that the base flow $\vec{U}$ verifies an equation of the generic form
\begin{equation}
\label{eq:fixed-point}
\begin{cases}
\displaystyle \vec{U}\cdot\grad{\vec{U}} + \grad{P} - \frac{1}{Re} \lapl{\vec{U}} - \diver{\parq{\nu_t\parton{\grad{\vec{U}} + (\grad{\vec{U}})^T}}} - \vec{f}_b = \vec{g},  \\[1ex]
\displaystyle \diver{\vec{U}}=0.
\end{cases}
\end{equation}
with some forcing term $\vec{g}$ to be determined. Such a forcing term is necessary since $\vec{U}$ is not a steady state solution to the unforced Navier-Stokes equations. This forcing term $\vec{g}$ differs from the forcing term $\vec{f}$ in equation \eqref{eq:eddy-resolvent} because equation \eqref{eq:eddy-resolvent} is linear with respect to $\vec{u}_s$ whereas equation \eqref{eq:fixed-point} implicitly contains the nonlinear term $A_s^2\vec{u}_s\cdot\grad{\vec{u}_s}$, with $A_s$ not vanishing. $\vec{g}$ can be computed explicitly using a nonlinear time-stepping algorithm as detailed in Appendix \ref{sec:appendix_forcing}. \\

Once $\vec{g}$ is known, the following nonlinear system is fully specified by
\begin{equation}
\label{eq:sns}
\begin{cases}
\displaystyle \derpar{\vec{u}}{t} + \vec{u}\cdot\grad{\vec{u}} + \grad{p} - \frac{1}{Re} \lapl{\vec{u}} - \diver{\parq{\nu_t\parton{\grad{\vec{u}} + (\grad{\vec{u}})^T}}} - \vec{f}_b = \vec{g},  \\[1ex]
\displaystyle \diver{\vec{u}}=0.
\end{cases}
\end{equation}
By construction, the base flow $\vec{U}$ is a steady state solution of this system. Moreover, when the amplitude of the streaks contained in $\vec{U}$ is sufficiently small, the forcing $\vec{g}$ converges towards the optimal harmonic forcing obtained from resolvent analysis $\vec{f}$, as verified in Appendix \ref{sec:appendix_forcing}. \\

Since  $\vec{U}$ is now a steady solution of equation \eqref{eq:sns}, LSA can be used rigorously. By perturbing the base flow $\vec{U}$ with a small perturbation $\vec{u}^{\prime\prime}$ such that $\vec{u}=\vec{U}+\vec{u}^{\prime\prime}$ and neglecting quadratic perturbation terms, one gets the following linear system for $\vec{u}^{\prime\prime}$ and the corresponding pressure perturbation $p^{\prime\prime}$,
\begin{equation}
\label{eq:slsa}
\begin{cases}
\displaystyle \derpar{\vec{u}^{\prime\prime}}{t} = -\vec{U}\cdot\grad{\vec{u}^{\prime\prime}} - \vec{u}^{\prime\prime}\cdot\grad{\vec{U}} - \grad{p^{\prime\prime}} + \frac{1}{Re} \lapl{\vec{u}^{\prime\prime}} + \diver{\parq{\nu_t\parton{\grad{\vec{u}^{\prime\prime}} + (\grad{\vec{u}^{\prime\prime}})^T}}},  \\[1ex]
\displaystyle \diver{\vec{u}}^{\prime\prime}=0.
\end{cases}
\end{equation}
The forcing term $\vec{g}$ cancels out in the secondary perturbation equation because it is constant, {\it i.e.} it depends only on $\vec{U}$ and not on $\vec{u}^{\prime\prime}$. Similarly, $\vec{f}_b$ cancels out because we assume that $\vec{u}^{\prime\prime}$ does not modify the flow rate.
Again, one can either choose to include the eddy viscosity term or to neglect it. We will compare the two approaches, consistently with the analysis in Section \ref{sec:resolvent}. Note that two different frozen eddy viscosity models, provided they predict exactly the same mean flow $\vec{U}$ and the same viscosity $\nu_t$, are expected to yield the same stability results for $\vec{u}^{\prime\prime}$.\\

Modal stability analysis assumes the following perturbation ansatz ($\iota$ denotes the imaginary unit):
\begin{equation}
\label{eq:ansatz}
    \displaystyle \vec{u}^{\prime\prime} = \tilde{\vec{u}}(y,z) e^{\sigma t + \iota k_x x} + \text{complex conjugate}, \quad \sigma \in \mathbb{C} \text{ and } k_x \in \mathbb{R},
\end{equation}
and an equivalent one for the pressure $p^{\prime\prime}$. 
Inserting the ansatz \eqref{eq:ansatz} into equation \eqref{eq:slsa}, the following eigenvalue problem is obtained:
\begin{equation}
\label{eq:stab}
\begin{cases}
\displaystyle \sigma \tilde{u} = - \iota k_xU\tilde{u} - V\derpar{\tilde{u}}{y} - W\derpar{\tilde{u}}{z} - \tilde{v}\derpar{U}{y} - \tilde{w}\derpar{U}{z} - \iota k_x\tilde{p} + \parton{\frac{1}{Re}+\nu_t}\lapl{\tilde{u}} + \frac{d\nu_t}{dy}\parton{\derpar{\tilde{u}}{y} + \iota k_x\tilde{v}},  \\[2ex]
\displaystyle \sigma \tilde{v} = - \iota k_xU\tilde{v} - V\derpar{\tilde{v}}{y} - W\derpar{\tilde{v}}{z} - \tilde{v}\derpar{V}{y} - \tilde{w}\derpar{V}{z} - \derpar{\tilde{p}}{y} + \parton{\frac{1}{Re}+\nu_t}\lapl{\tilde{v}} + 2\frac{d\nu_t}{dy}\derpar{\tilde{v}}{y},  \\[2ex]
\displaystyle \sigma \tilde{w} = - \iota k_xU\tilde{w} - V\derpar{\tilde{w}}{y} - W\derpar{\tilde{w}}{z} - \tilde{v}\derpar{W}{y} - \tilde{w}\derpar{W}{z} - \derpar{\tilde{p}}{z} + \parton{\frac{1}{Re}+\nu_t}\lapl{\tilde{w}} + \frac{d\nu_t}{dy}\parton{\derpar{\tilde{w}}{y} + \derpar{\tilde{v}}{z}},  \\[2ex]
0 = \displaystyle \iota k_x \tilde{u} + \derpar{\tilde{v}}{y} + \derpar{\tilde{w}}{z},
\end{cases}
\end{equation}
with $\lapl{}=-k_x^2 + \partial^2_y + \partial^2_z$.\\

The secondary stability problem is solved in a two-dimensional $y-z$ domain using Fourier collocation in the spanwise direction and Chebyshev collocation in the wall-normal direction. \\

\subsection{Exploiting the spanwise periodicity of the streaks}
\label{sec:block_circulant}

Standard LSA is usually performed by assuming eigenvectors with a wavelength matching the fundamental wavelength of the base flow. This refers here to the spanwise wavelength $\lambda_{z,s}$ of the streaks. By constrast, the wavelength can be freely chosen in the $x$-direction. In that case, the linearised system \eqref{eq:stab}, once discretised, leads to a generalised eigenvalue problem of the form \begin{equation}
    \sigma \mathsfbi{B}_0\mathsfbi{\tilde{q}}_0 = \mathsfbi{A}_0\mathsfbi{\tilde{q}}_0.
    \label{eq:eig0}
\end{equation}
with $\mathsfbi{\tilde{q}}_0$ a vector of size $M$, containing the four primitive variables $\tilde{u}$, $\tilde{v}$, $\tilde{w}$ and $\tilde{p}$ for each point of the computational grid, and $\mathsfbi{A}_0$ and $\mathsfbi{B}_0$ matrices of size $M\times M$. The matrix $\mathsfbi{B}_0$ is the so-called prolongation matrix that arises when using a velocity-pressure formulation \citep{cerqueira2014eigenvalue}.\\

We are interested in this study in predicting instabilities of the base flow occuring over wavelengths larger than that of the base flow. This can be achieved by simply considering a numerical domain containing several concatenated copies of the spatially periodic base flow. However, solving equation \eqref{eq:eig0} becomes computationally more costly as the domain size increases. This cost can be reduced exploiting the $z$-periodicity of the base flow by following the study of \citet{schmid2017stability}. We therefore consider a given (integer) number $N_u$ of base flow units concatenated in the $z$-direction, such that the total spanwise size of the domain used for LSA is now $N_u\lambda_{z,s}$. After spatial discretisation, the stability problem \eqref{eq:stab} becomes a new generalised eigenvalue problem of the form 
\begin{equation}
    \sigma \mathsfbi{B}\mathsfbi{\tilde{q}} = \mathsfbi{A}\mathsfbi{\tilde{q}},
    \label{eq:eig}
\end{equation}
where $\mathsfbi{\tilde{q}}$ has $N_u$ times the size of the original $\mathsfbi{\tilde{q}}_0$ vector in \eqref{eq:eig0}, i.e. $MN_u$, and the matrices $A$ and $B$ have now size $MN_u \times MN_u$. Owing to the periodicity of the base flow and to the assumed periodicity of the eigenvector over $N_u$ unit cells, the matrix $\mathsfbi{A}$ has a block-circulant structure. One can hence diagonalize it into a block-diagonal matrix $\mathsfbi{\hat{A}}=\mathsfbi{P^H}\mathsfbi{A}\mathsfbi{P}$ with $N_u$ blocks $\mathsfbi{\hat{A}}_0,\mathsfbi{\hat{A}}_1,...,\mathsfbi{\hat{A}_{N_u}}$, thereby reducing one large eigenvalue problem of size $N_u M \times N_uM$ down to $N_u$ manageable eigenproblems, each of size $M \times M$. 
The matrix $\mathsfbi{P}\in\mathbb{C}^{N_uM\times N_uM}$ is given by the Kronecker product $\mathsfbi{J}\otimes\mathsfbi{I}_{M}$ with $\mathsfbi{J}\in\mathbb{C}^{N_u\times N_u}$, $J_{j+1,k+1}=\rho_j^k/\sqrt{N_u}$ and $\mathsfbi{I}_M$ the identity matrix of size $M\times M$ \citep{schmid2017stability}.
Each of these spectral subproblems involves a matrix of size $M$, namely $\mathsfbi{\hat{A}}_j=\mathsfbi{A}_0+\rho_j\mathsfbi{A}_1 + \rho_j^2\mathsfbi{A}_2 + ... + \rho_j^{N_u-1}\mathsfbi{A}_{N_u-1}$ for each $j=0,...,N_u-1$, featuring $\mathsfbi{A}_0,\mathsfbi{A}_1,...$ which are the blocks of $\mathsfbi{A}$. The general eigensolutions of \eqref{eq:eig} can be written \citep{schmid2017stability} as 
\begin{equation}
\mathsfbi{\tilde{q}}=\left(\mathsfbi{\tilde{q}}_j,\rho_j\mathsfbi{\tilde{q}}_j,...,\rho_j^{N_u-1}\mathsfbi{\tilde{q}}_j\right)^T,
\label{eq:q}
\end{equation}
based on the known eigenvectors $\mathsfbi{\tilde{q}}_j$ of the matrices $\mathsfbi{\hat{A}}_j$.\\

In \eqref{eq:q}, the complex numbers $\rho_j$, $j=1,\dots,N_u$, are the $N_u$ different roots of unity solutions of the equation $\rho^{N_u}=1$. This defining property for the $\rho_j$'s ensures that the generalised eigenvector $\mathsfbi{\tilde{q}}$ is periodic in $z$ over the distance $N_u\lambda_{z,s}$. This allows one to use spectral methods based on periodic Fourier functions. Each $\rho_j$ has modulus one and can be rewritten as $\rho_j=\exp(2\pi \gamma_j \iota)$. The $\gamma$'s are defined in $[0,1)$. They take only discrete values $0,1/N_u,2/N_u,\dots$ These exponents are ideally interpreted, for a given eigenvector, as the ratio between the wavelength of the base flow and that of the eigenvector. In the particular case where $\gamma=0$ ($\rho_1=1$), the eigenvector has the same periodicity as the base flow, i.e. $N_u=1$. One recovers thus the classical LSA results, since  $\rho_1=1$ and $\gamma=0$. In the special case where $\gamma$ is of the form $1/Q$ with $Q$ a non-zero integer, then the fundamental wavelength of the eigenvector is exactly $Q\lambda_{z,s}$, and it can be labelled  \textit{sub-harmonic}. In the general case, for finite $N_u$, only rational values of $\gamma$ can be tackled using this method and irrational values are excluded. The physical range of meaningful values of $\gamma$ is continuous and involves irrational values as well. In practice, it is advised to consider a value of $N_u$ as large as possible in order to have access to a well-discretised range of values of $\gamma$ in $[0,1)$. As a consequence, $\gamma$ can be interpreted as a \textit{detuning factor} \citep{jouin2024}. Note moreover that the symmetry $z \leftarrow -z$ inherent to the base flow, implies that $\gamma$ and $1-\gamma$ are associated with the same eigenvalue and eigenvector. This restricts the study of the eigenspectrum to the range of values $\gamma \in [0,0.5]$. Finally, the union of the spectra of the sub-matrices $\mathsfbi{\hat{A}}_j$ yields the spectrum of the original matrix $\mathsfbi{A}$. In practice, the latter is found by looping over $j=0,...,N_u-1$, which is indirectly a loop over the values of $\gamma$ in $[0,0.5]$. This is the basis of the computation shown in Figure \ref{fig:eigenspectra}.\\

The exponent $\gamma$ plays the same role of the Floquet modulation parameter in spatial Floquet theory \citep{herbert1988secondary}, despite the fact that the block-circulant matrix method and spatial Floquet analysis (also called Bloch formalism) are formally different. However, \textit{a posteriori} the two methods give identical results \citep{jouin2024}.\\

The main results of the paper are obtained using $N_u=50$ base flow units. Each unit was discretised with $65$ points in the wall-normal direction and $20$ collocation points in the spanwise direction. The number of points in the spanwise direction refers to the discretisation of a single streak wavelength. Therefore, it turns out to be sufficient for the relatively narrow spanwise width of 100 wall units considered here. The convergence of the results with the number of collocation points and with the number of base flow units is assessed in Appendix \ref{sec:appendix_convergence}. The code was validated in a previous study \citep{ciola2024large}.

\subsection{Critical Reynolds number}
\label{sec:critical_Re}

The first question addressed in this work is whether a critical Reynolds number exists below which spatial modulations appear. In the work of \citet{kashyap2024linear}, based on a linearisation around the asymptotic mean flow ({\it e.g.} without streaks), no such critical threshold was identified. It is well established that streaks become unstable for sufficiently large amplitude \citep{schoppa2002coherent}. However, classical streak instability is characterized by wavelengths which are either identical ($\gamma=1$) or double ($\gamma=0.5$) of the wavelength of the streak \citep{andersson2001breakdown}, whereas the focus of this work is on instabilities that modulate the streaky flow over much larger scales. Therefore, the second point to be addressed is whether the unstable mode at criticality features an important large-scale component. In this subsection we focus specifically on $k_x=0.18$, the critical streamwise wavenumber reported by \citet{kashyap2022linear}.\\

\begin{figure}
\centering
\raisebox{1.in}{(a)}\includegraphics[width=0.45\textwidth]{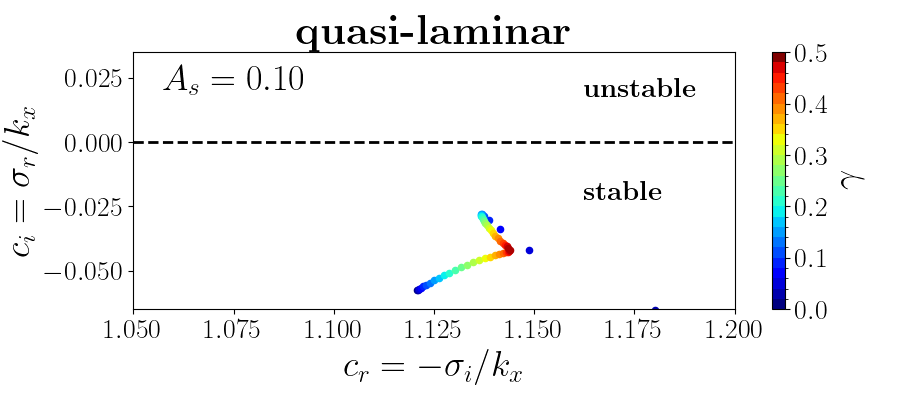}\quad
\raisebox{1.in}{(b)}\includegraphics[width=0.45\textwidth]{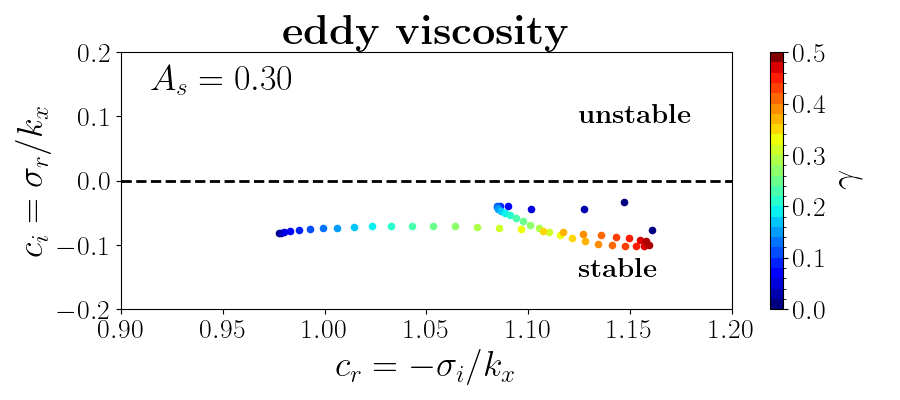} \\
\raisebox{1.in}{(c)}\includegraphics[width=0.45\textwidth]{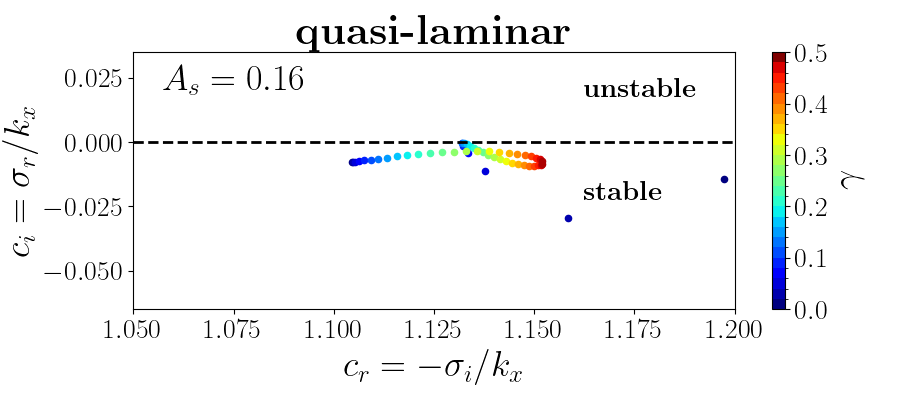}\quad
\raisebox{1.in}{(d)}\includegraphics[width=0.45\textwidth]{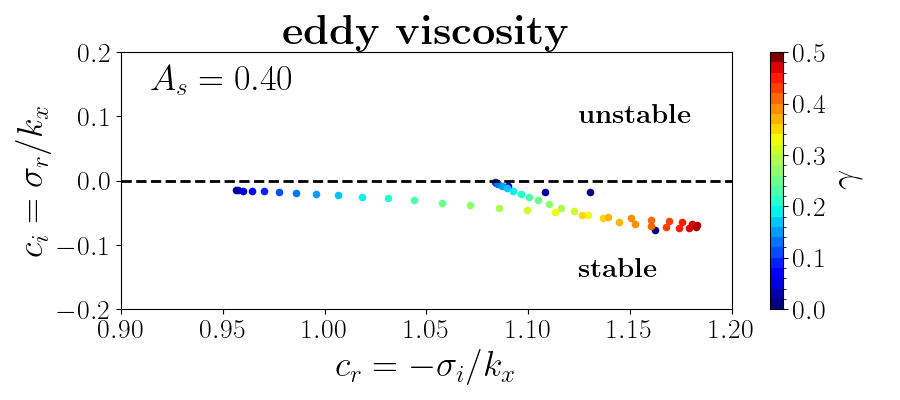}\\
\raisebox{1.in}{(e)}\includegraphics[width=0.45\textwidth]{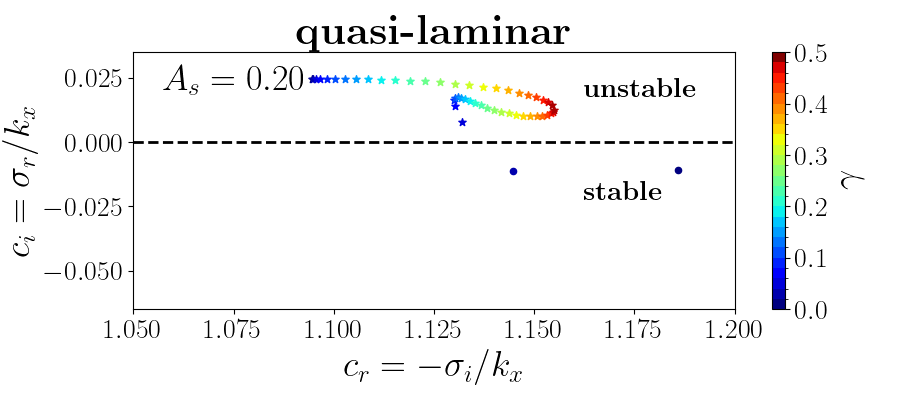}\quad
\raisebox{1.in}{(f)}\includegraphics[width=0.45\textwidth]{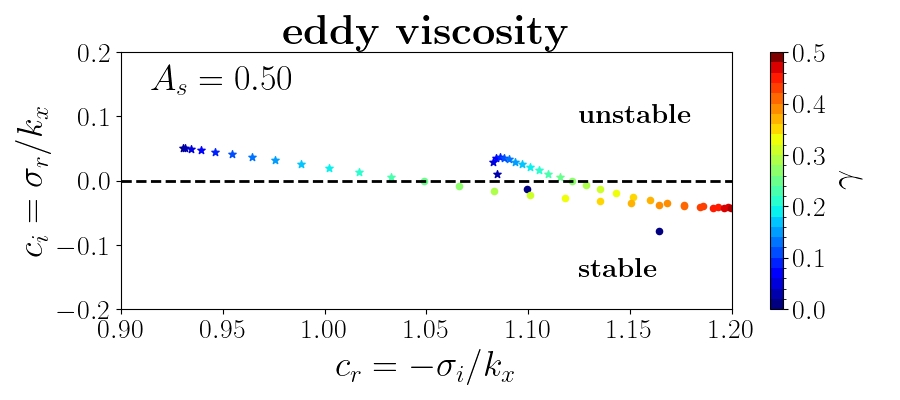}
\caption{Eigenvalues for the streak stability problem (only a subset of the computed spectra is shown) for $Re_{\tau}=71$ and $k_x=0.18$. The eigenvalues ($\bullet$ for stable modes, $\star$ for unstable modes) are coloured with the corresponding root of unity factor $\gamma = j/N_u$ for $j=0,\dots,N_u/2$ (the eigenvalues for $\gamma \in (0.5,1.0)$ are equal to those for $\gamma \in (0,0.5)$ and the corresponding modes are the same up to a reflection in the spanwise direction). (a-c-e) Quasi-laminar model; (b-d-f) eddy viscosity model. Streak amplitudes increase from top to bottom and are (a) $0.10$, (c) $0.16$ and (e) $0.20$ for the quasi-laminar model and (b) $0.30$, (d) $0.40$ and (f) $0.50$ for the eddy viscosity model.}
\label{fig:eigenspectra}
\end{figure}

The effect of the streak amplitude $A_s$ on the eigenvalues associated with \eqref{eq:eig} is documented in figure \ref{fig:eigenspectra}. This figure shows a close-up view in the complex plane of the leading branch of eigenvalues, {\it i.e.} the least stable (or the most unstable) one. It shows the eigenvalues in the form of complex phase velocities $c=-\sigma/\iota k_x$. The figure displays only a small part of the computed eigenspectrum. There is no instability elsewhere in the range of parameters investigated. As explained in the previous section, these spectra are the union of $N_u$ sub-spectra, each associated with a different root of unity. Therefore, in figure \ref{fig:eigenspectra} the eigenvalues are coloured according to the respective factor $\gamma$, which identifies the relevant root of unity and takes values in $[0,1)$. In practice, only the range $[0,0.5]$ needs to be shown since the eigenmodes corresponding to $\gamma$ in $(0.5,1)$ are obtained from those in $(0,0.5)$ by spanwise reflection  (see subsection \ref{sec:block_circulant}). Focusing on $Re_{\tau}=71$ and $k_x=0.18$, the figure shows that the branch becomes unstable as the $A_s$ is increased. The critical amplitude for the quasi-laminar model (left panels) is $\approx 0.16$, whereas for the eddy viscosity model it is larger, $\approx 0.4$ (right panels). After inspection of the eigenvalues for several $A_s$, we report that, for both models, the first mode that becomes unstable at $k_x=0.18$ has $\gamma > 0$. \\

\begin{figure}
\centering
\raisebox{1.4in}{(a)}\includegraphics[width=0.45\textwidth]{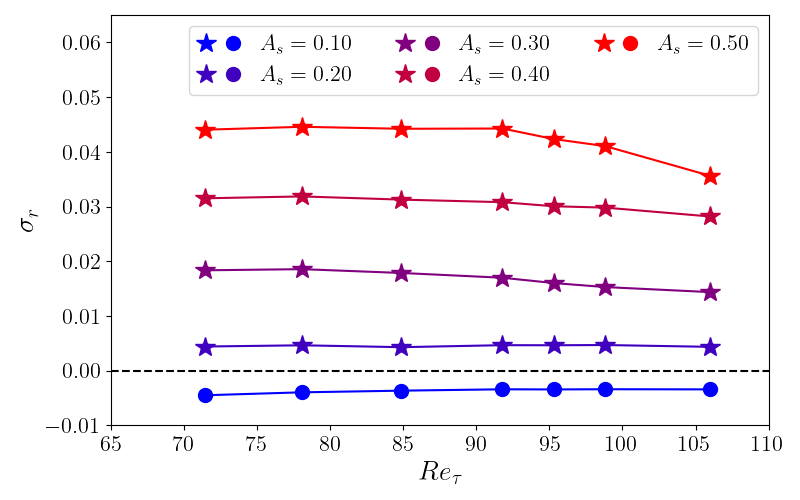}\quad
\raisebox{1.4in}{(b)}\includegraphics[width=0.45\textwidth]{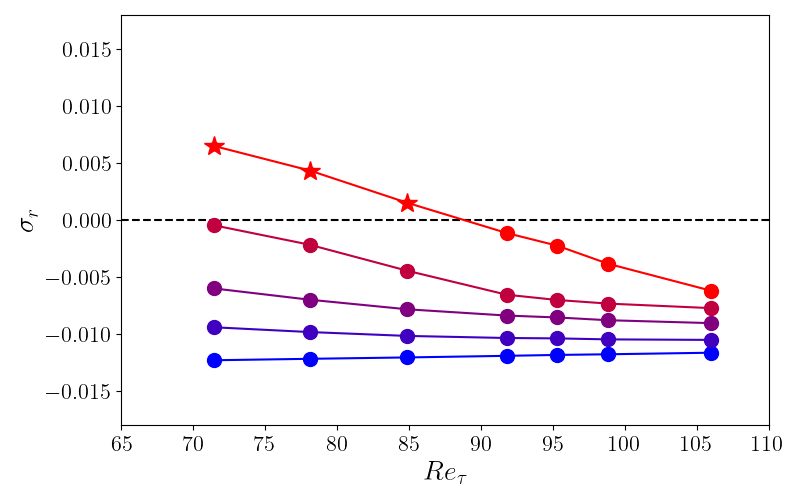} \\
\raisebox{1.4in}{(c)}\includegraphics[width=0.45\textwidth]{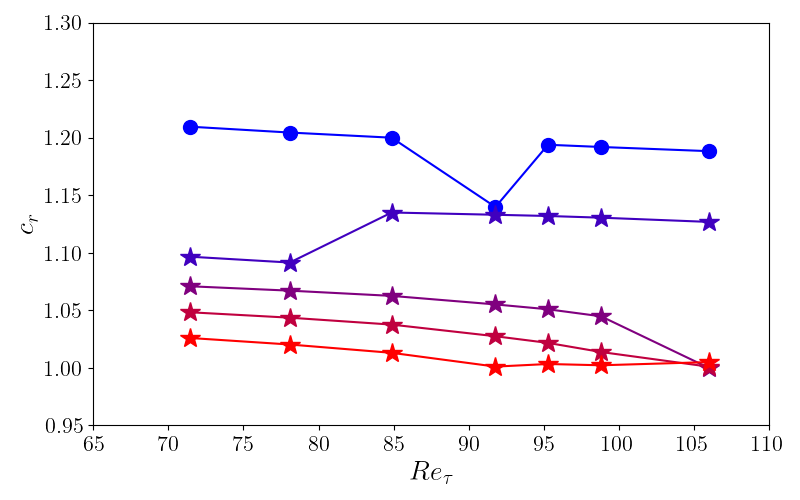}\quad
\raisebox{1.4in}{(d)}\includegraphics[width=0.45\textwidth]{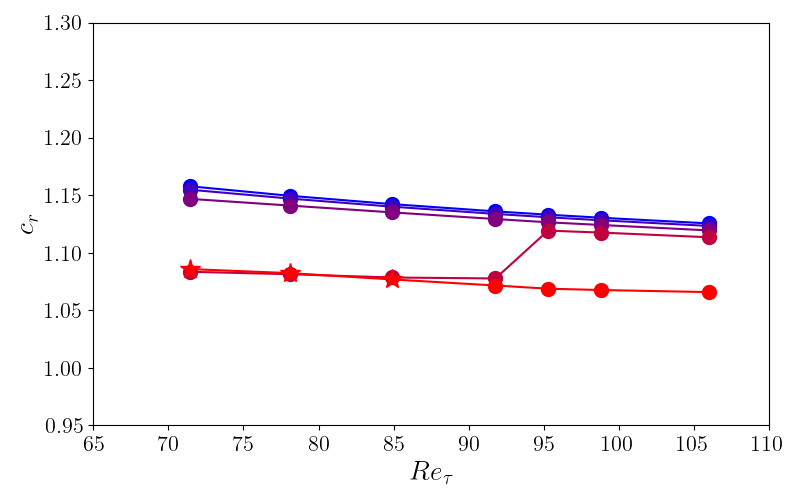}
\caption{Variation of the leading eigenvalues with $Re$ and streak amplitude $A_s$ for $k_x=0.18$ ($\bullet$ for stable modes, $\star$ for unstable modes). (a-b): growth rate. (c-d): phase velocity. (a-c) Quasi-laminar model; (b-d) eddy viscosity model.}
\label{fig:gr_vs_Re}
\end{figure}

The $Re_{\tau}$-dependence of the leading eigenvalue is documented in figure \ref{fig:gr_vs_Re} for varying $A_s$. The first row shows the leading growth rates for the two models considered, respectively quasi-laminar or using an eddy viscosity. In the quasi-laminar model the growth rates become positive without any apparent dependence on $Re$. A critical amplitude can still be defined, independently of $Re_\tau$, but no critical Reynolds number exists. For the eddy viscosity model, the growth rate increases with $A_s$ and depends also on the value of $Re_{\tau}$. As a result, for $A_s$ large enough, as $Re_{\tau}$ is decreased, the growth rate of the most unstable mode becomes positive at a well defined critical Reynolds number. A striking point is the qualitative difference between the results of the two models. The presence of a critical $Re$ with the eddy viscosity model is a qualitative improvement with respect to the mean flow analysis in \cite{kashyap2024linear}.\\

The bottom panels of figure \ref{fig:gr_vs_Re} show the phase velocity of the leading modes with respect to $Re_\tau$ and $A_s$. The values globally overestimate by roughly $5-10\%$ the advection velocities of turbulent bands measured from DNS \citep{tuckerman2014turbulent}, however they display a similar decrease with $Re$. \\

Besides the fact that the critical value itself depends on the value of $A_s$, another point of concern is the fact that the instability and the Reynolds number dependence are observed only for large amplitudes. The issue whether these amplitudes are realistic will be addressed in the discussion section.

\subsection{Unstable (large-scale) modes}

\begin{figure}
\centering
\raisebox{1.in}{(a)}\includegraphics[width=0.45\textwidth]{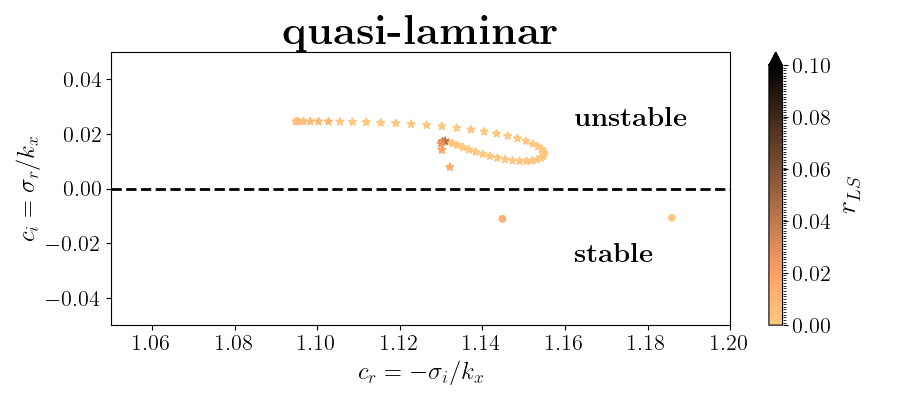}\quad
\raisebox{1.in}{(b)}\includegraphics[width=0.45\textwidth]{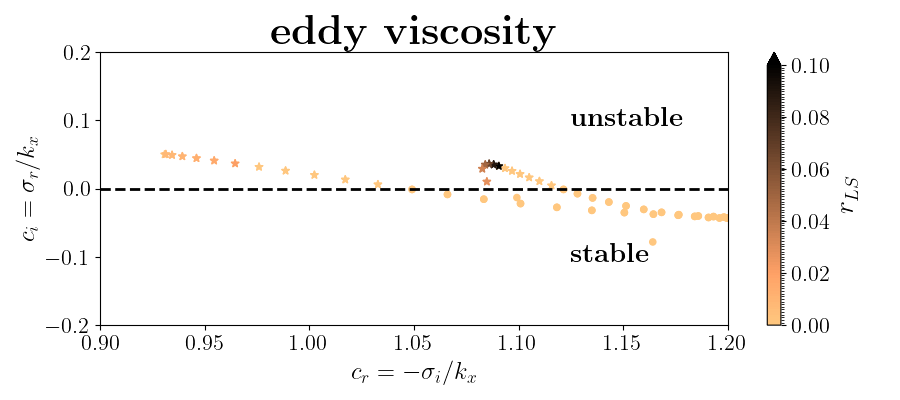}
\caption{Eigenvalues coloured by the large-scale spanwise energy ratio $r_{LS}$ (only a subset of the computed spectra is shown) for $Re_{\tau}=71$ and $k_x=0.18$ ($\bullet$ for stable modes, $\star$ for unstable modes). (a) Quasi-laminar model and $A_s=0.2$; (b) eddy viscosity model and $A_s=0.5$.}
\label{fig:eigenspectra_lsr}
\end{figure}

The above results are encouraging and call for an examination of the unstable eigenmodes. Notably, no spanwise wavelength of the modes is explicitly imposed in the {\it ansatz} in \eqref{eq:ansatz}. All detuned eigenmodes contain energy in wavelengths larger than the base flow wavelength $\lambda_{z,s}$, yet the question arises whether some eigenmodes have a particularly prominent large-scale component. In this respect, the detuning factor $\gamma$ yields only partial information. Therefore, an additional quantification of the large-scale property is introduced by
\begin{equation}
    r_{LS} = \dfrac{\bigint_0^2 \; \sum\limits_{\abs{k_z}<k_z^c} \abs{\check{\vec{u}}(y,k_z)}^2 \; dy}{\bigint_0^2 \; \sum\limits_{k_z} \abs{\check{\vec{u}}(y,k_z)}^2 \; dy},
\end{equation}
where $\check{\vec{u}}$ denotes the Fourier transform of the mode in $z$ and $k_z^c$ is a cut-off wavenumber that represents a higher limit for a Fourier mode to be considered large-scale. The value $k_z^c=0.5$ was chosen after inspection of the DNS and it corresponds to a wavelength which is roughly $10$ times greater than the wavelength of the streaks depending on $Re$.
The new quantity $r_{LS}$ is the ratio of the eigenmode's energy at large wavelengths to the total eigenmode energy. It is used to color the eigenvalues in {\it e.g.} figure \ref{fig:eigenspectra_lsr}, where darker points correspond to eigenmodes with a pronounced large-scale character. Again, the qualitative difference between the quasi-laminar model and the eddy viscosity model is significant. For the eddy viscosity model the leading mode has a large-scale factor near $10\%$, to be compared with much weaker values for the quasi-laminar model. \\

\begin{figure}
\centering
\raisebox{0.7in}{(a)}\includegraphics[width=0.98\textwidth]{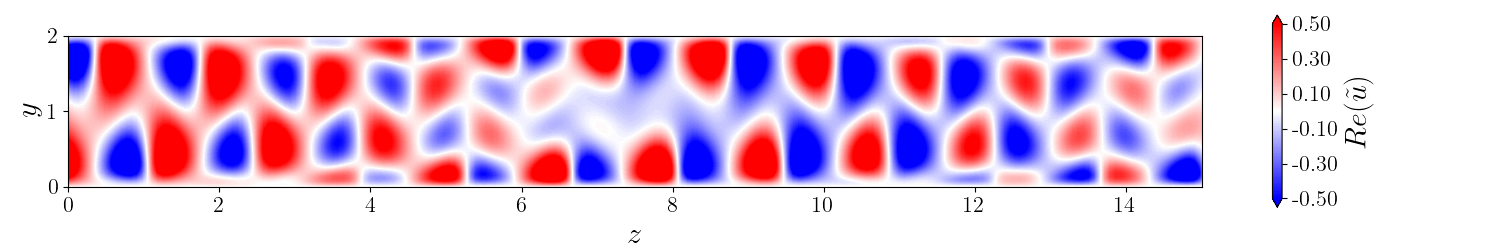}\\
\raisebox{0.7in}{(b)}\includegraphics[width=0.98\textwidth]{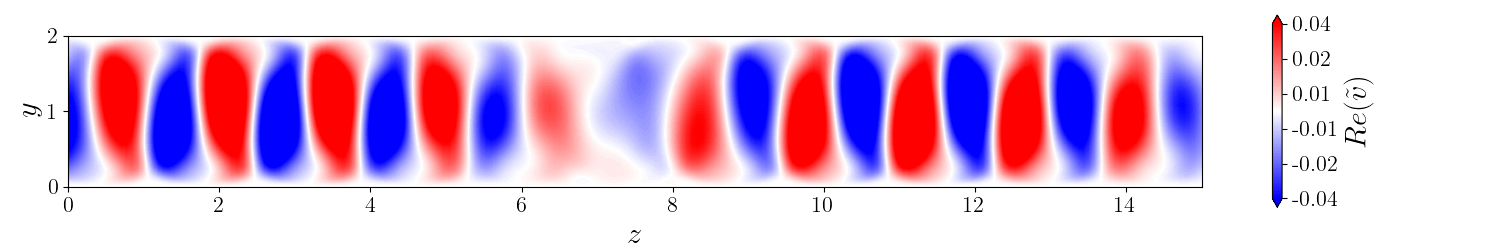}\\
\raisebox{0.7in}{(c)}\includegraphics[width=0.98\textwidth]{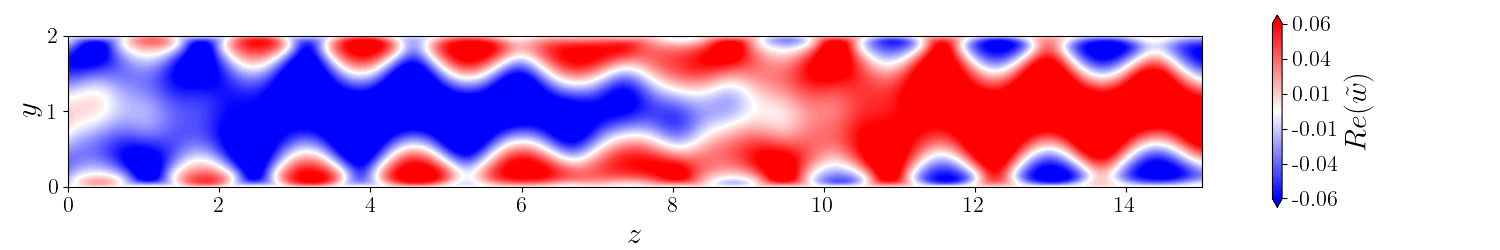}
\caption{Leading eigenmode for the eddy viscosity model with $Re_{\tau}=71$, $k_x=0.18$ and $A_s=0.50$ ($\lambda_{z,s}\approx1.41$ at this value of $Re_{\tau}$). The modes are normalised to $\max_{y,z} \abs{Re(\tilde{u})} = 1$. Real part of (a) streamwise velocity component, (b) wall-normal velocity component and (c) spanwise velocity component.}
\label{fig:eigenmode_yz}
\end{figure}

\begin{figure}
\centering
\raisebox{1.1in}{(a)}\includegraphics[width=0.3\textwidth]{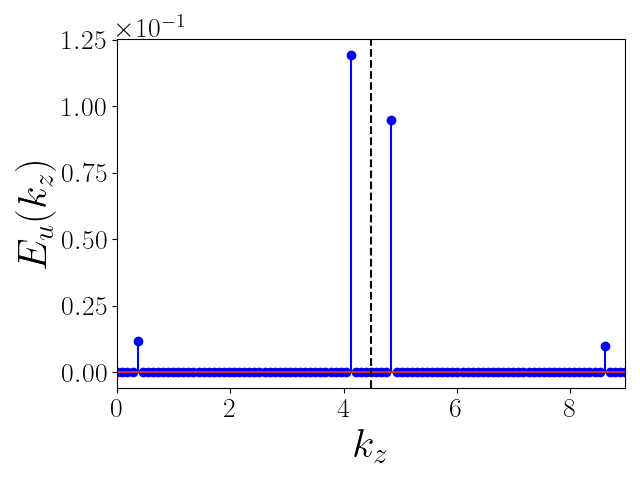}
\raisebox{1.1in}{(b)}\includegraphics[width=0.3\textwidth]{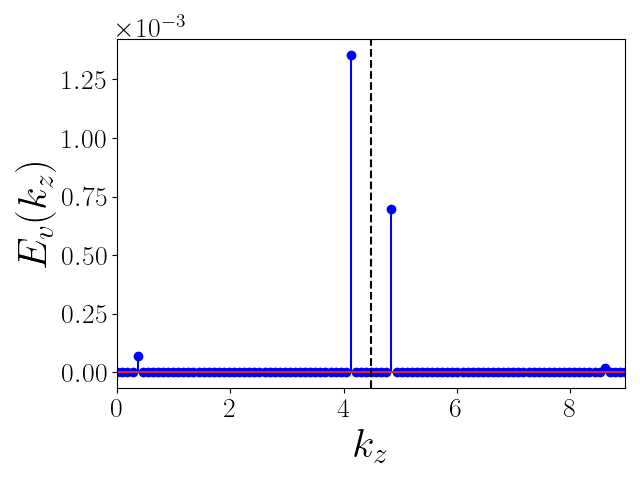}
\raisebox{1.1in}{(c)}\includegraphics[width=0.3\textwidth]{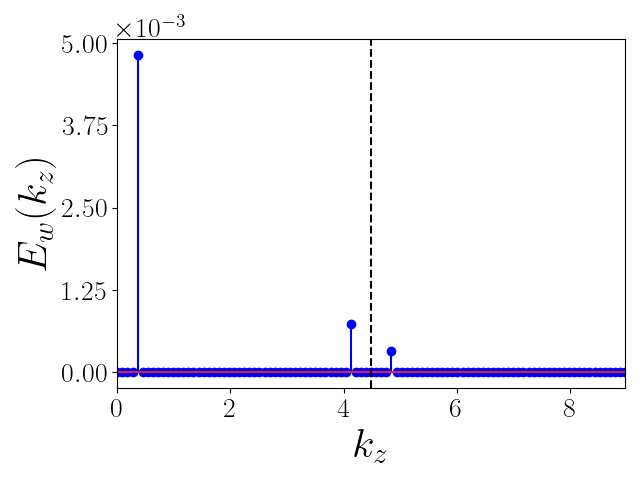}
\caption{Spanwise Fourier decomposition of the leading eigenmode for the same parameters as Fig. \ref{fig:eigenmode_yz}. The figure shows the wall-normal integrated squared Fourier coefficient of the real part of the (a) streamwise, (b) wall-normal and (c) spanwise velocity components. The black dashed line denotes the wavenumber of the base flow $2\pi/\lambda_{z,s}$.}
\label{fig:eigenmode_fourier_z}
\end{figure}

The  spatial structure of this interesting eigenmode for the eddy viscosity model is shown in figure \ref{fig:eigenmode_yz} in the $y-z$ plane. The eigenmode is characterised by a small-scale modulation with the same spanwise periodicity of the base flow streaks, and by a large-scale modulation. In particular, the large-scale modulation of the wall-normal velocity component appears as an amplitude modulation. On the spanwise component, it appears as a large-scale flow. As for the streamwise component, the large-scale character is less clear but the large-scale modulating feature is still clearly observed. 
These qualitative observations are confirmed looking at the Fourier decomposition of the modes along $z$. The real part of each eigenmode component $\hat{u}_i(y,z)$ is decomposed in Fourier series. The wall-normal integrated squared modulus of the Fourier coefficients gives energies as a function of spanwise wavenumber $E_i(k_z)$, which are plotted in figure \ref{fig:eigenmode_fourier_z} for the three velocity components. The figure shows that the eigenmode can be seen as a superposition of waves. The first wave has a small wavenumber and is responsible for the large-scale character of the eigenmode. We note that $r_{LS}$ was designed to give the ratio of the energy of this large-scale component with respect to the total energy of the eigenmode. The other waves have wavelengths near $\lambda_{z,s}$ (dashed line in the figure) or smaller. Moreover, figure \ref{fig:eigenmode_fourier_z} (c) shows that the large-scale wave is dominant in the spanwise velocity component, as suggested by figure \ref{fig:eigenmode_fourier_z} (c). Lastly, we note that the large-scale spanwise wavenumber of this mode is $k_z \approx 0.36$. Given that $k_x=0.18$ for this eigenmode, it means that the modulation forms an angle of $\tan^{-1}(k_x/k_z)\approx 26.6^{\circ}$ with respect to the streamwise direction. This angle is not far from the $23^{\circ}$ reported by \citet{kashyap2022linear} and the $22.5^{\circ}$ found by \citet{benavides2025model}.

\begin{figure}
\centering
\includegraphics[width=\textwidth]{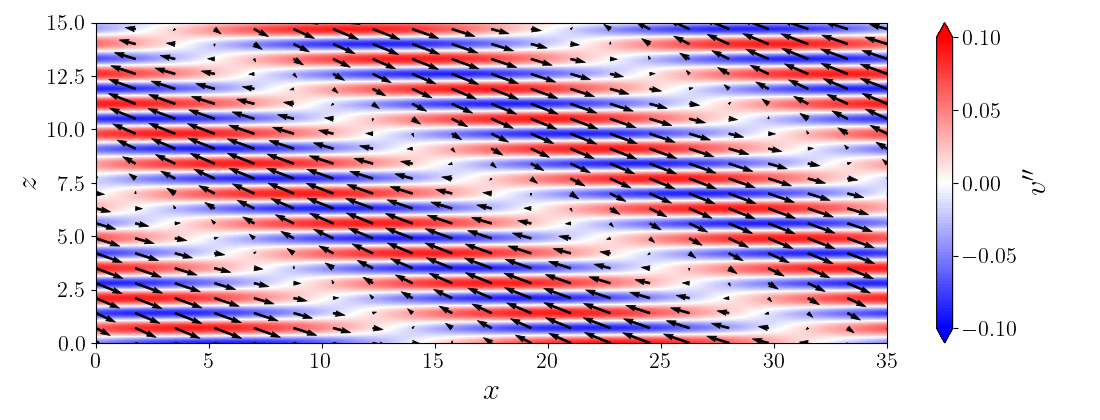}
\caption{Leading eigenmode (real part of the full ansatz) for the same parameters as Fig. \ref{fig:eigenmode_yz}. Wall-normal velocity component (shaded contours) at midplane and wall-parallel large-scale flow (black arrows) in the $x-z$ plane. The large-scale flow is obtained by integrating the wall-parallel velocity components in the wall-normal direction.}
\label{fig:eigenmode_xz}
\end{figure}

The above analysis is consistent with DNS observations interpreted as a modulation of the small scales together with a wall-parallel large-scale flow. This ideal picture is confirmed by figure \ref{fig:eigenmode_xz}, which shows the eigenmode integrated along the wall-normal direction $y$ in the horizontal plane $x-z$. The arrows portray the streamwise and spanwise velocity components forming the large-scale flow. The displayed velocity field is fully consistent with most observations in the patterning regime \citep{duguet2013oblique,tuckerman2014turbulent}. Similar observations also apply to other unstable eigenmodes that belong to the branch and are detuned. There are also unstable modes which do not show any large-scale modulation, related to classical sinuous/varicose streaks instabilities which are involved in the self-sustaining cycle \citep{hamilton1995regeneration,waleffe1997self}. Their presence is expected from the classical literature on streak instabilities and they are hence not the focus of this work. 

\begin{figure}
\centering
\raisebox{1.4in}{(a)}\includegraphics[width=0.45\textwidth]{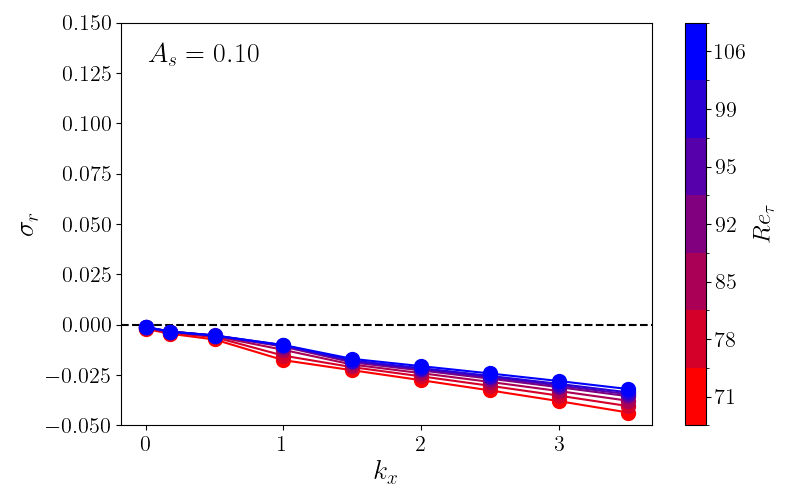}\quad
\raisebox{1.4in}{(b)}\includegraphics[width=0.45\textwidth]{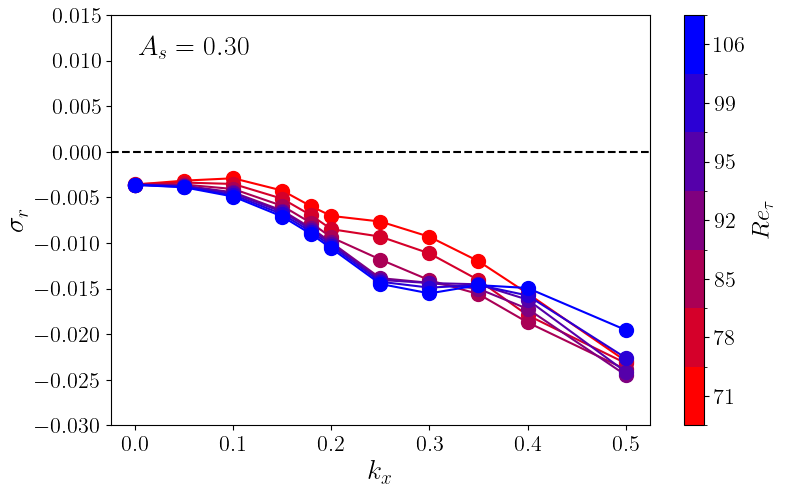} \\
\raisebox{1.4in}{(c)}\includegraphics[width=0.45\textwidth]{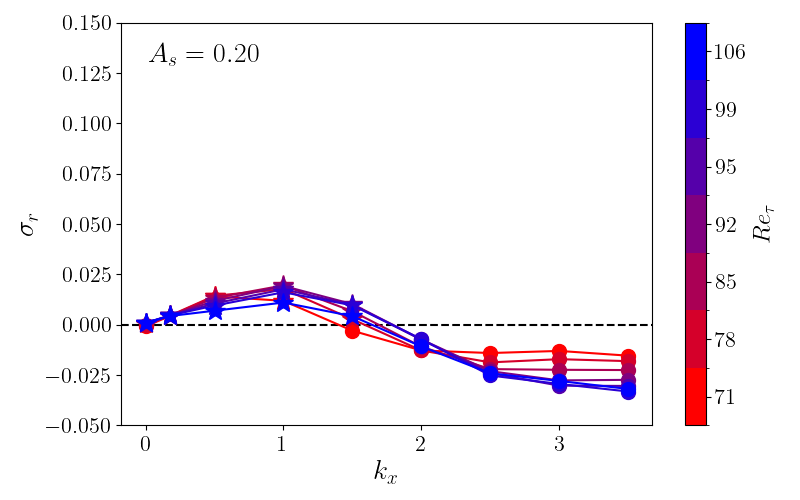}\quad
\raisebox{1.4in}{(d)}\includegraphics[width=0.45\textwidth]{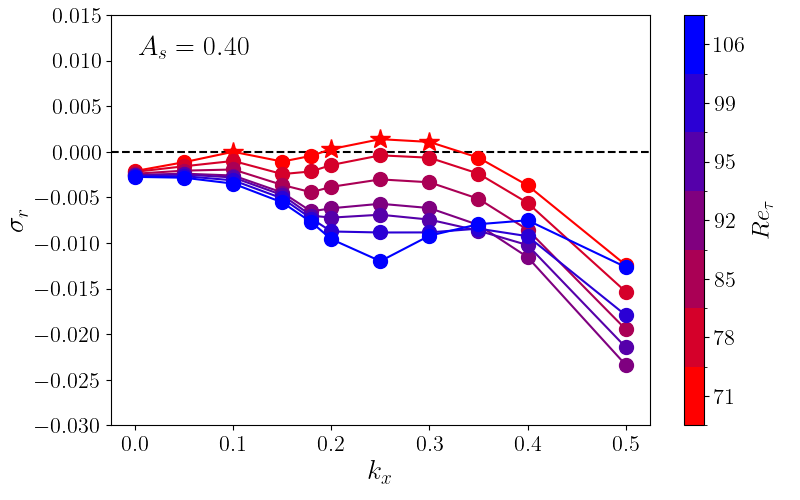}\\
\raisebox{1.4in}{(e)}\includegraphics[width=0.45\textwidth]{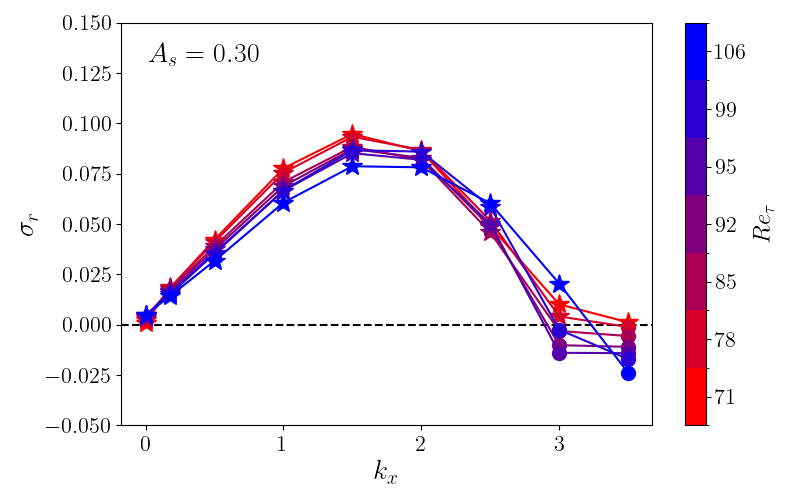}\quad
\raisebox{1.4in}{(f)}\includegraphics[width=0.45\textwidth]{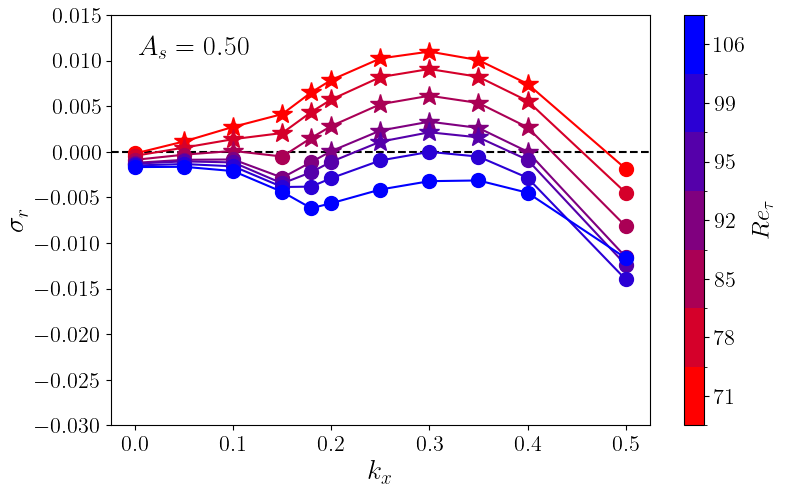}
\caption{Variation of the leading growth rate $\sigma_r$ with $Re$ and streamwise wavenumber $k_x$ ($\bullet$ for stable modes, $\star$ for unstable modes). (a-c-e) Quasi-laminar model; (b-d-f) eddy viscosity model. Note the different scale in $k_x$ between left and right panels. Streak amplitudes $A_s$ increase from top to bottom and are (a) $0.10$, (c) $0.20$ and (e) $0.30$ for the no-closure cases and (b) $0.30$, (d) $0.40$ and (f) $0.50$ for the cases with eddy viscosity.}
\label{fig:gr_vs_kx}
\end{figure}

When the streamwise wavenumber is fixed to a value consistent with the value from \citet{kashyap2022linear}, namely $k_x=0.18$, the eddy viscosity model gives a group of eigenmodes characterised by a large-scale modulation that becomes unstable as $Re_{\tau}$ is lowered. It is now necessary to assess which streamwise wavenumber is selected by the system. Figure \ref{fig:gr_vs_kx} shows the dependence of the leading growth rate on the streamwise wavenumber for several values of $A_s$ and for all $Re$ under consideration. For low $A_s$, the growth rate monotonically decreases with $k_x$ both for the quasi-laminar model and for the eddy viscosity model. The same behaviour was reported in \cite{kashyap2024linear} for the stability analysis of the mean flow, {\it i.e.} the case $A_s=0$. As the streak amplitude is increased, the growth rate is maximal for a non-zero streamwise wavenumber. For the quasi-laminar model, the maximum is unique and the curves for the different $Re_{\tau}$ overlap. Moreover, the maximum occurs for $k_x \approx 1$, which implies $\lambda_x \approx 3-6$. These wavelengths are too short to be representative of the large-scale patterns of interest. Instead, with the eddy viscosity model, the instability is limited to $k_x \approx \mathcal{O}(0.1)$, {\it i.e.} $\lambda_x \approx 30-40$, which is consistent with expectations. In conclusion, only when the Reynolds stresses are taken into account does stability analysis select a relevant large-scale streamwise wavenumber.\\

\subsection{Critical Amplitude and Wavelengths}

This subsection is devoted to a deeper analysis of the eddy viscosity model at critical conditions. The critical parameters are defined as the parameters for which the leading mode has zero growth rate. Usually in linear stability studies, as in \citet{kashyap2022linear}, a critical $Re$ and critical wavenumbers are reported. In our model, we have a degree of freedom given by the streak amplitude. One can either define a critical $Re$ for a given $A_s$ (note that the critical $Re$ is defined only for sufficiently high $A_s$, figure \ref{fig:gr_vs_Re} (b)) or define a critical $A_s$ for each $Re$, so that a neutral curve $Re-A_s$ can be computed. For each point on this neutral curve there is a critical eigenmode with a streamwise wavenumber $k_x^c$. Moreover, a critical spanwise wavenumber $k_z^c$ can also be defined since the eigenmodes are characterised by one single large-scale wavelength (see figure \ref{fig:eigenmode_fourier_z}).

\begin{figure}
\centering
\raisebox{1.1in}{(a)}\includegraphics[width=0.3\textwidth]{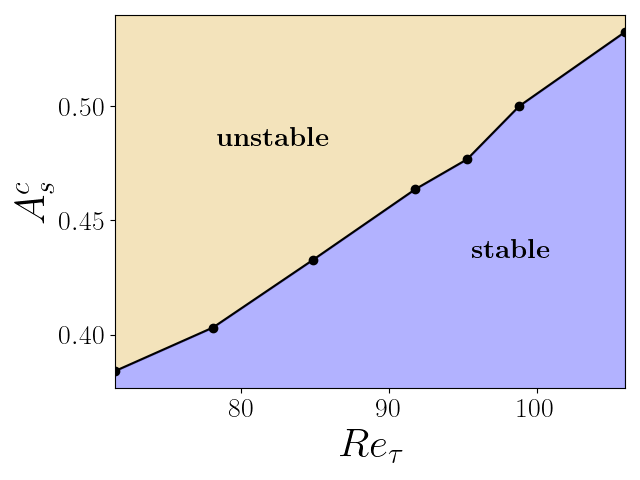}
\raisebox{1.1in}{(b)}\includegraphics[width=0.3\textwidth]{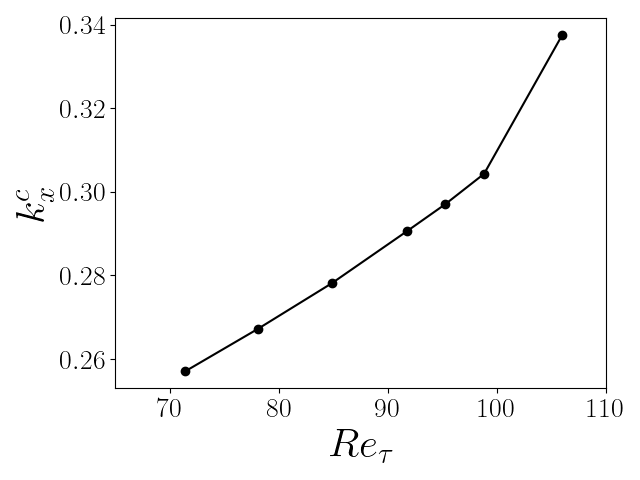}
\raisebox{1.1in}{(c)}\includegraphics[width=0.3\textwidth]{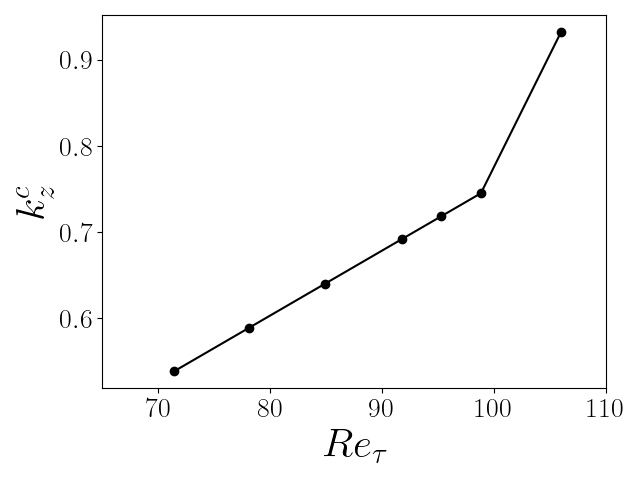}
\caption{Critical $A_s$, $k_x$ and large-scale $k_z$ as a function of $Re_{\tau}$ for the eddy viscosity model.}
\label{fig:critical_params}
\end{figure}

The neutral curve is computed for the considered value of $Re_{\tau}$ by coupling a line search algorithm in $k_x$ with a bisection algorithm for $A_s$. In the inner loop we look for the $k_x$ giving maximum growth rate for a given $A_s$. Then, $A_s$ is bisected until the absolute value of the maximum growth rate falls below a tolerance of $10^{-5}$. The result is presented in figure \ref{fig:critical_params} (a). It can be seen that the critical amplitude grows almost linearly with $Re$. This means that for fixed $A_s$ the instability is obtained by decreasing $Re$, as explained in Subsection \ref{sec:critical_Re} and consistently with \citet{kashyap2022linear}. The neutral curve has not been computed for the quasi-laminar model since the results presented in figure \ref{fig:gr_vs_Re} already suggest that the critical amplitude is independent of $Re$ in the considered range. 

Panels (b) and (c) of figure \ref{fig:critical_params} show the critical wavenumbers. They increase with $Re$, indicating that the instability tends towards smaller wavelengths at high $Re$. This is also consistent with DNS observations \citep{kashyap2020flow}. At $Re_{\tau}=95$, we obtain $k_x^c\approx0.3$ and $k_z^c\approx 0.71$, which are slightly greater than the $k_x^c\approx0.18$ and $k_z^c\approx0.42$ of \citet{kashyap2022linear}. However, the ratio between the wavenumbers, hence the angle of the eigenmode with the streamwise direction, is approximately the same ($\approx 23^{\circ}$). We report that our critical angle slowly decreases with $Re$ from $25.5^{\circ}$ at $Re_{\tau}=71$ to $20.0^{\circ}$ at $Re_{\tau}=106$ (not shown).

\section{Energy budget analysis}
\label{sec:eba}

In order to interpret physically the results of the stability analysis of the streaky base flow, an equation equivalent to the Reynolds-Orr equation for the most unstable eigenvector is derived in this section \citep{schoppa2002coherent, schmid2002stability,albensoeder2001,alizard2015linear}. For a simpler derivation an index notation is introduced, where $x_1$, $x_2$ and $x_3$ correspond, respectively, to $x$, $y$ and $z$ and $u_1$, $u_2$ and $u_3$ to $u$, $v$ and $w$. Summation over repeated indices is implicit, and $i$ takes the values $1$, $2$ and $3$, whereas $j$ takes only the values $2$ and $3$. Lastly, in this section an asterisk denotes the complex conjugate. 

\subsection{Derivation \label{sec:eba_derivation}}

By taking the scalar product of $\tilde{\vec{u}}^*$ with the momentum equation in \eqref{eq:stab}, one obtains the following scalar equation
\begin{equation}
\label{eq:localOR}
\begin{aligned}
    \sigma \tilde{u}_i^*\tilde{u}_i = & -\iota k_x U \tilde{u}_i^*\tilde{u}_i - U_j\tilde{u}_i^*\derpar{\tilde{u}_i}{x_j} - \tilde{u}_i^*\tilde{u}_j\derpar{U_i}{x_j} -\iota k_x\tilde{u}^*\tilde{p} - \tilde{u}_j^*\derpar{\tilde{p}}{x_j} - k_x^2 \parton{\frac{1}{Re} + \nu_t} \tilde{u}_i^*\tilde{u}_i \\[1ex]
    & + \parton{\frac{1}{Re} + \nu_t}\tilde{u}_i^*\derparm{\tilde{u}_i}{x_j}{x_j} + \tilde{u}_i^*\frac{d\nu_t}{dy}\derpar{\tilde{u}_i}{y} + \iota k_x \tilde{u}^*\tilde{v}\frac{d\nu_t}{dy} + \tilde{u}_j^*\frac{d\nu_t}{dy}\derpar{\tilde{v}}{x_j}.
\end{aligned}
\end{equation}
Moreover, the following identities are considered:
\begin{align}
    -\iota k_x\tilde{u}^*\tilde{p} - \tilde{u}_j^*\derpar{\tilde{p}}{x_j} & = - \derpar{}{x_j}\parton{\tilde{u}_j^*\tilde{p}} \qquad \text{(using the continuity equation)}; \\[1ex]
    \frac{1}{Re}\tilde{u}_i^*\derparm{\tilde{u}_i}{x_j}{x_j} & = \frac{1}{Re}\derpar{}{x_j}\parton{\tilde{u}_i^*\derpar{\tilde{u}_i}{x_j}} - \frac{1}{Re}\derpar{\tilde{u}_i^*}{x_j}\derpar{\tilde{u}_i}{x_j}; \\[1ex]
    \nu_t\tilde{u}_i^*\derparm{\tilde{u}_i}{x_j}{x_j} & = \derpar{}{x_j}\parton{\nu_t\tilde{u}_i^*\derpar{\tilde{u}_i}{x_j}} - \nu_t\derpar{\tilde{u}_i^*}{x_j}\derpar{\tilde{u}_i}{x_j} - \tilde{u}_i^*\frac{d\nu_t}{dy}\derpar{\tilde{u}_i}{y}.
\end{align}

Substituting the above identities in equation \eqref{eq:localOR}, then integrating on the secondary-stability $y-z$ domain ($\Omega$) and normalizing the eigenmode such that $\int_{\Omega}\tilde{u}_i^*\tilde{u}_i \; d\Omega = 1$, a decomposition for the complex eigenvalue $\sigma$ is obtained :
\begin{equation}
\label{eq:globalOR}
    \sigma = T_a + \Prod - \Diss - \Diss^c + C,
\end{equation}
where
{\allowdisplaybreaks
\begin{align}
    T_a = &   \underbrace{-\iota k_x \int_{\Omega} U \tilde{u}_i^*\tilde{u}_i \; d\Omega}_{\displaystyle T_{a1}} \underbrace{- \int_{\Omega} U_j\tilde{u}_i^*\derpar{\tilde{u}_i}{x_j} \; d\Omega}_{\displaystyle T_{a2}}; \\[1ex]
    \Prod \; = &  \underbrace{-\int_{\Omega} \tilde{u}^*\tilde{v}\derpar{U}{y} \; d\Omega}_{\displaystyle \Prod_{uy}} \underbrace{-\int_{\Omega} \tilde{u}^*\tilde{w}\derpar{U}{z} \; d\Omega}_{\displaystyle \Prod_{uz}} \underbrace{-\int_{\Omega} \tilde{v}^*\tilde{v}\derpar{V}{y} \; d\Omega}_{\displaystyle \Prod_{vy}} \underbrace{-\int_{\Omega} \tilde{v}^*\tilde{w}\derpar{V}{z} \; d\Omega}_{\displaystyle \Prod_{vz}} \\[1ex]
    & \; \underbrace{-\int_{\Omega} \tilde{w}^*\tilde{v}\derpar{W}{y} \; d\Omega}_{\displaystyle \Prod_{wy}} \underbrace{-\int_{\Omega} \tilde{w}^*\tilde{w}\derpar{W}{z} \; d\Omega}_{\displaystyle \Prod_{wz}} ; \\[1ex]
    \Diss \; = & \; \frac{k_x^2}{Re}\int_{\Omega}\tilde{u}_i^*\tilde{u}_i \; d\Omega + \frac{1}{Re}\int_{\Omega} \derpar{\tilde{u}_i^*}{x_j}\derpar{\tilde{u}_i}{x_j} \; d\Omega; \\[1ex]
    \Diss^c = & \; k_x^2\int_{\Omega}\nu_t\tilde{u}_i^*\tilde{u}_i \; d\Omega + \int_{\Omega} \nu_t\derpar{\tilde{u}_i^*}{x_j}\derpar{\tilde{u}_i}{x_j} \; d\Omega;\\[1ex]
    \label{eq:Cterm}
    C \hspace{0.15cm} = & \; \underbrace{\iota k_x \int_{\Omega} \tilde{u}^*\tilde{v} \frac{d\nu_t}{dy} \; d\Omega}_{\displaystyle C_1} + \underbrace{\int_{\Omega} \frac{d\nu_t}{dy}\tilde{u}_j^*\derpar{\tilde{v}}{x_j} \; d\Omega}_{\displaystyle C_2}.
\end{align}
}

The equation \eqref{eq:globalOR} represents a Reynolds-Orr type energy budget for an eigenvector perturbation to the two-dimensional base flow, extended to the turbulent regime modelled by a turbulent eddy viscosity.
We should note that the following terms integrate to zero because of the boundary conditions in $y$ and $z$:
\begin{align}
    T_p = & \; -\int_{\Omega}\derpar{}{x_j}\parton{\tilde{u}_j^*\tilde{p}} \; d\Omega \equiv 0; \\[1ex]
    T_d = & \; \frac{1}{Re} \int_{\Omega}\derpar{}{x_j}\parton{\tilde{u}_i^*\derpar{\tilde{u}_i}{x_j}} \; d\Omega \equiv 0; \\[1ex]
    T_d^c = & \; \int_{\Omega} \derpar{}{x_j}\parton{\nu_t\tilde{u}_i^*\derpar{\tilde{u}_i}{x_j}} \; d\Omega \equiv 0.
\end{align}

The relation \eqref{eq:globalOR} shows that the complex-valued eigenvalue $\sigma$ associated with one given eigenmode results from several contributions: advective transport ($T_a$), production due to base flow gradients ($\Prod$), dissipation due to molecular viscosity ($\Diss$), dissipation due to eddy viscosity ($\Diss^c$) and production due to eddy viscosity gradients ($C$). Of course the last two terms are present only when the eddy viscosity model is used. Some remarks need to be made about these quantities: (i) $T_a$ is purely imaginary, hence it contributes to the angular frequency $Im(\sigma)$ but does not contribute to the growth rate $Re(\sigma)$ (as expected for a transport term); (ii) $\Diss$, $\Diss^c$ and $C$ are purely real with $\Diss \geq 0$ and $\Diss^c \geq 0$ and thus contribute to the damping of the mode, whereas $C$ generally contributes positively to the growth rate; (iii) $\Prod$ is complex but contributes mainly to the real part of the eigenvalue with a positive amount. The main contribution to the imaginary part of $\sigma$ (hence to the phase velocity of the mode) comes from $T_{a1}$, {\it i.e.} from the mean flow advection.\\

\begin{figure}
\centering
\raisebox{1.5in}{(a)}\includegraphics[width=0.45\textwidth]{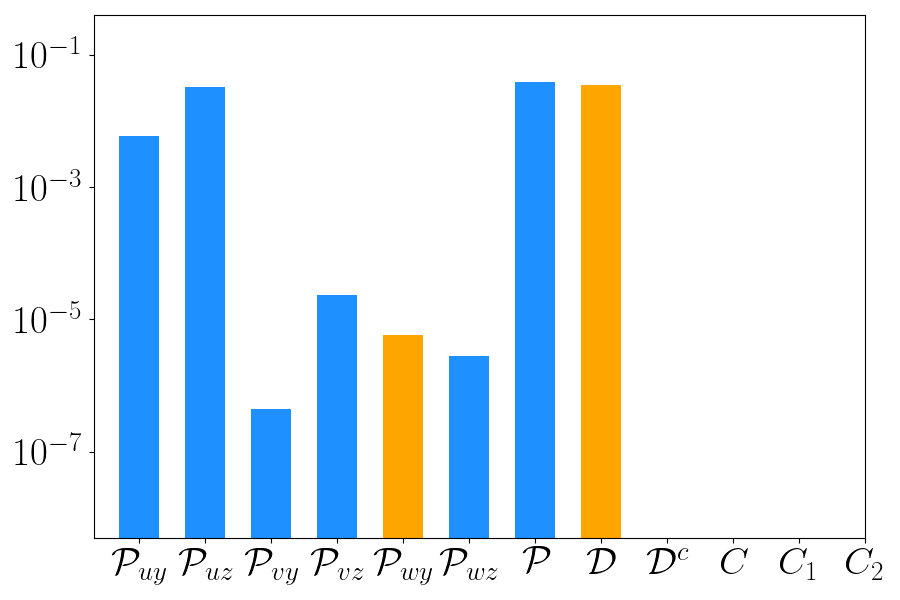}\quad
\raisebox{1.5in}{(b)}\includegraphics[width=0.45\textwidth]{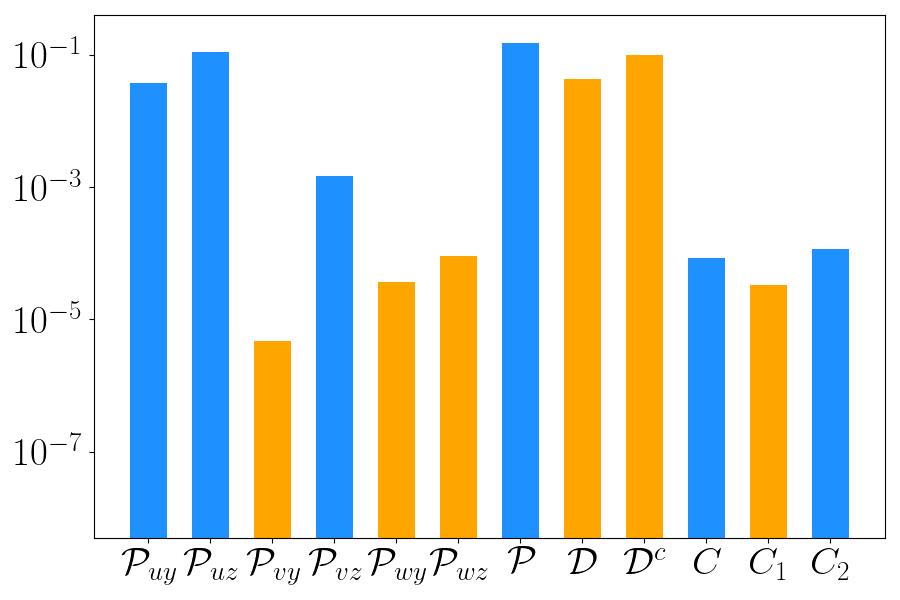}
\caption{Energy budget for the leading eigenmode at $Re_{\tau}=71$ and $k_x=0.18$ for (a) the quasi-laminar model ($A_s=0.2$) and (b) the eddy viscosity model ($A_s=0.5$). Blue bars denote positive contributions to the growth rate while orange bars denote negative contributions. For the meaning of the labels see equations (\ref{eq:globalOR}-\ref{eq:Cterm}) in the text.}
\label{fig:eb_bars}
\end{figure}

The terms contributing to the real part of the eigenvalues (the modal growth rate) are analysed in more detail on specific examples. We focus on the parameter values $Re_{\tau}=71$, $k_x=0.18$, $A_s=0.2$ for the quasi-laminar model and $A_s=0.5$ for the eddy viscosity model. Figure \ref{fig:eb_bars} shows the various terms from equation \eqref{eq:globalOR} for the two cases. For the quasi-laminar case, the important terms are $\mathcal{P}_{uy}$, $\mathcal{P}_{uz}$ and $\mathcal{D}$. The remaining production terms are several order of magnitude smaller. For the eddy viscosity case, the eddy viscosity dissipation $\mathcal{D}^c$ is another important term. On the contrary, the production due to eddy viscosity gradients $C$ appears as negligible. Nonetheless, this does not imply that the wall-normal variation of the eddy viscosity is not important. All the quantities plotted are integrals that involve the eigenmodes and, in general, it can be expected that the wall-normal variation of $\nu_t$ contributes to the structure of the eigenmodes. It was verified by inspection that similar considerations apply to all the cases considered in the following analysis.

\subsection{Results \label{sec:eba_results} }

\begin{figure}
\centering
\raisebox{1.4in}{(a)}\includegraphics[width=0.45\textwidth]{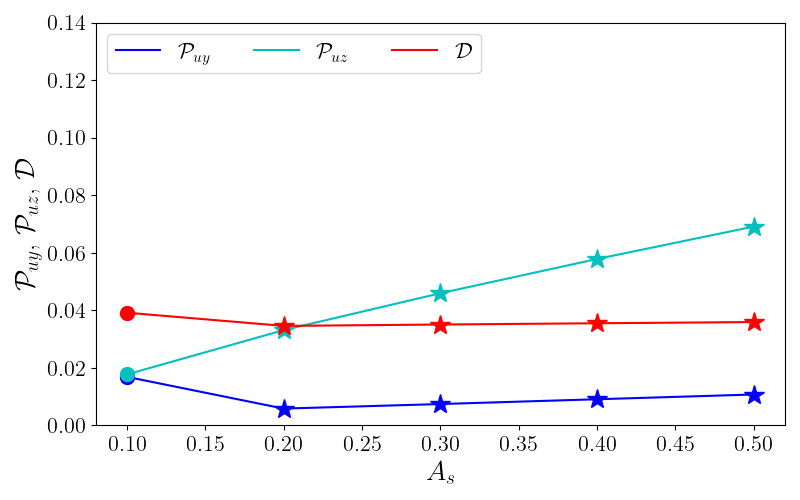}\quad
\raisebox{1.4in}{(b)}\includegraphics[width=0.45\textwidth]{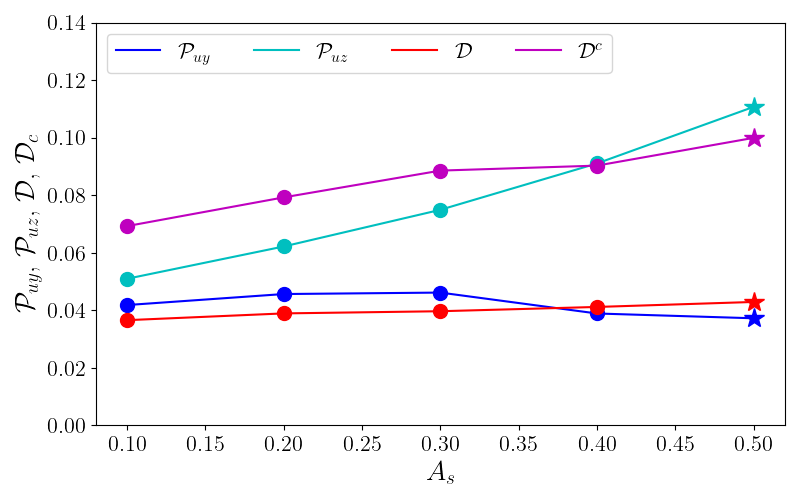}\\
\raisebox{1.4in}{(c)}\includegraphics[width=0.45\textwidth]{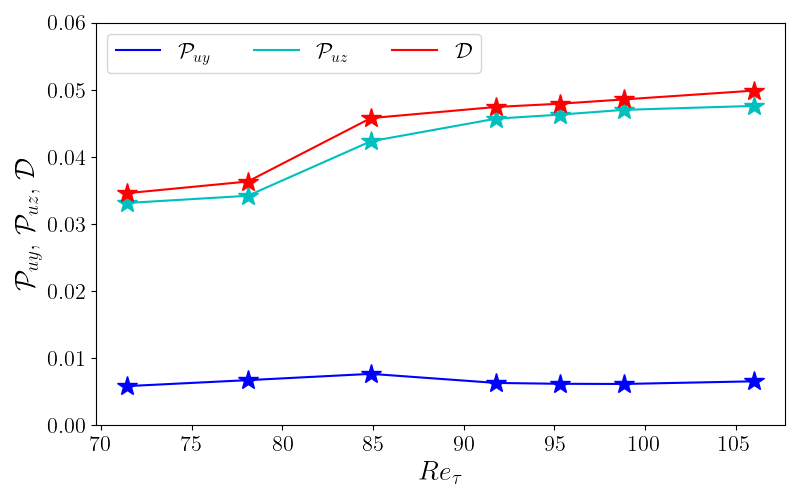}\quad
\raisebox{1.4in}{(d)}\includegraphics[width=0.45\textwidth]{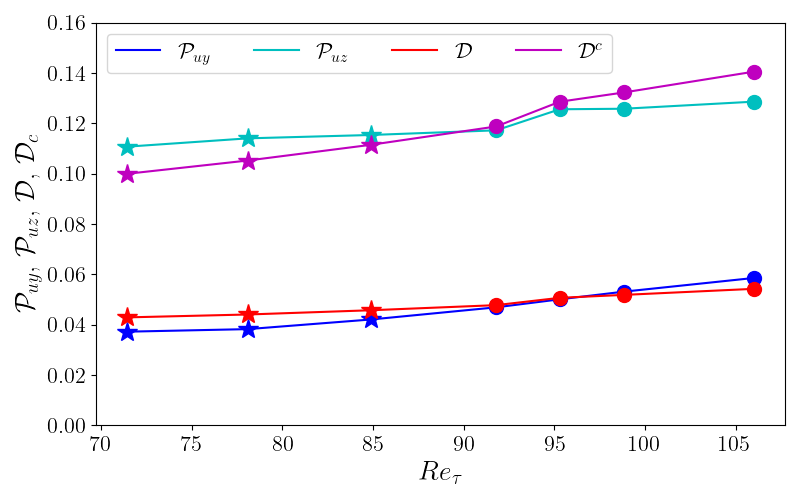}
\caption{Energy budget for the leading eigenmode with $k_x=0.18$. (a-b) Variation with streak amplitude $A_s$ for fixed $Re_{\tau}=71$. (c-d) Variation with $Re$ for fixed streak amplitude $A_s=0.2$ in (c) and $A_s=0.5$ in (d). (a-c): Quasi-laminar model; (b-d) eddy viscosity model. For the meaning of the labels see equations (\ref{eq:globalOR}-\ref{eq:Cterm}) in the text.}
\label{fig:eb_vars}
\end{figure}

Having identified the dominant terms, their dependence on the stability parameters is analyzed in figure \ref{fig:eb_vars}. Concerning the effect of $A_s$ (panels (a) and (b)), for both models the production term inducing instability is the one associated with spanwise gradients of the base flow, $\mathcal{P}_{uz}$. For the eddy viscosity model, a larger amplitude is needed to compensate the dissipation due to the eddy viscosity.\\

The panels (c) and (d) in figure \ref{fig:eb_vars} illustrate why a critical $Re$ is observed only within the eddy viscosity model. For the quasi-laminar model (panel (c)) production and dissipation follow essentially the same trend, resulting in a growth rate almost independent of $Re_{\tau}$. Conversely, for the eddy viscosity model (panel (d)) the eddy viscosity dissipation increases faster than the production when the $Re_{\tau}$ is increased. Consequently, at some critical value of $Re_{\tau}$ the eddy viscosity dissipation takes over the production and stabilizes the eigenmode. This suggests a $Re$-dependence for the growth rates encoded in the $Re$-dependence of the eddy viscosity, an interesting result pointing at the role of the turbulent fluctuations in this instability.\\

\begin{figure}
\centering
\raisebox{1.4in}{(a)}\includegraphics[width=0.45\textwidth]{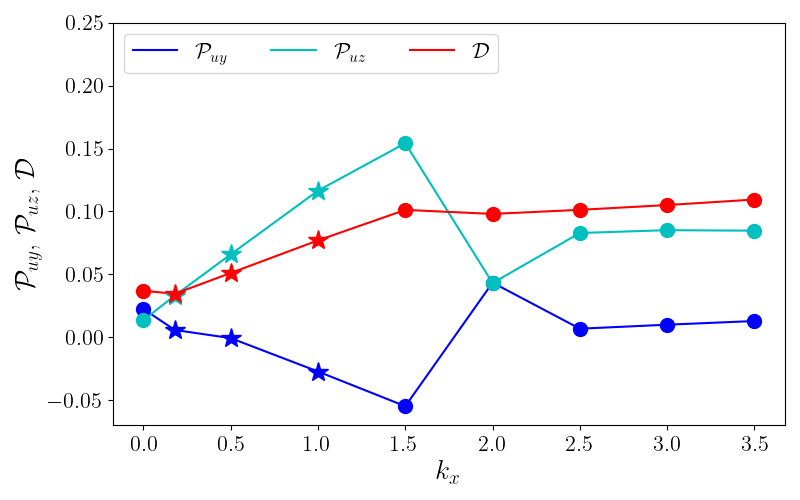}\quad
\raisebox{1.4in}{(b)}\includegraphics[width=0.45\textwidth]{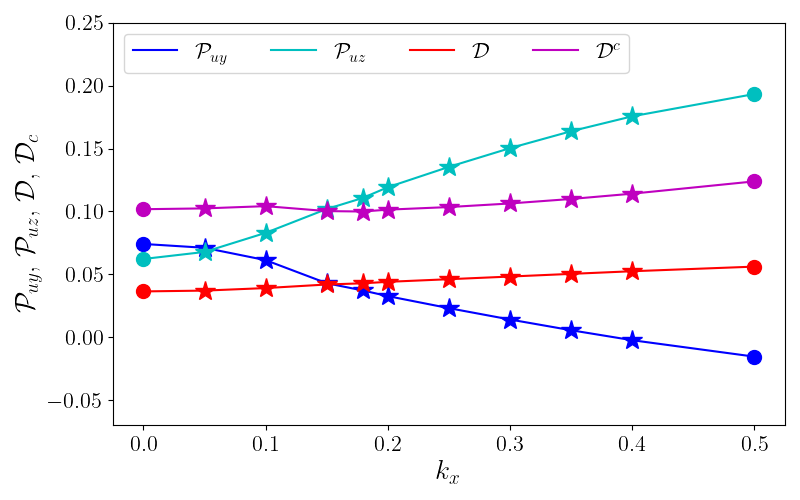}\\
\caption{Energy budget for the leading eigenmode as $k_x$ is varied. $Re_{\tau}=71$. (a) Quasi-laminar model with $A_s=0.2$; (b) eddy viscosity model with $A_s=0.5$. Note the different scale of the abscissa in the two panels. For the meaning of the labels see equations (\ref{eq:globalOR}-\ref{eq:Cterm}) in the text.}
\label{fig:eb_vars_kx}
\end{figure}

The two panels of figure \ref{fig:eb_vars_kx} illustrate how the energy budget changes with the streamwise wavenumber $k_x$. For both models $\mathcal{P}_{uy}$ and $\mathcal{P}_{uz}$ depart from each other, the former decreasing while the latter increases. For the quasi-laminar model in panel (a), this behaviour changes drastically beyond $k_x=1.5$, indicating a change in the nature of the leading eigenmode. However, the main difference between the two models is again the presence of the eddy viscosity dissipation that stabilizes the modes for $k_x > 0.4$. \\

In conclusion, the energy budget analysis shows that the inclusion of the eddy viscosity induces the observed parameter dependence mainly via the additional damping term $\mathcal{D}^c$.\\ 

\section{Discussion}

The streak field selected in this study corresponds to the most amplified structure, computed using resolvent analysis after linearising around the mean flow. An eigenvalue problem was formulated by linearising around a synthetic base flow, constructed as the mean flow, to which the optimal streaks have been added with a tunable amplitude. It was solved numerically by allowing for detuned eigenmodes, {\it i.e.} spatially subharmonic modes whose wavelength is not a simple multiple of the spanwise wavelength of the streaks. Visual comparison between the most unstable large-scale eigenmode and the structure of laminar-turbulent patterns, as observed in 
numerical simulations, suggests that this instability is indeed relevant to the appearance of the pattern. 

The starting hypothesis for this approach differs greatly from that of \citet{liu2021structured}, who have considered linearisation around the analytical laminar base flow solution. The streak instability considered here takes place in a turbulent environment whose effect needs to be modelled. In fact, this work shows that the turbulent fluctuations contribute to the sought instability in two ways. First, the coherent part of the fluctuations (the streaks) make the flow unstable to various wavelengths. Secondly, the turbulent fluctuations, via the eddy viscosity term, damp high wavelengths selecting specific large-scale modes, thereby introducing a cut-off in wavenumber. As a result of this competition, large-scale modes can be destabilised at low enough $Re$ provided $A_s$ is large enough. This result highlights the need to correctly model unresolved turbulent motions in linear studies of turbulence in agreement with other recent works \citep{illingworth2018,morra2019relevance,symon2023use}. The present model is an improvement compared to the approach in \citet{kashyap2024linear}, which featured the eddy viscosity in the governing equations but not the streak mode in the base flow.\\

Another important assumption of the model is the high degree of symmetry of the optimal streaks obtained from resolvent analysis. By construction the resolvent modes are harmonic in $z$ and uniform in $x$. As a consequence of the harmonicity in $z$ they do not modify the mean flow. In particular, the typical $100$ wall units spacing of the streaks was chosen as the spanwise wavelength for the resolvent analysis. In appendix \ref{sec:appendix_3Dstreaks}, it is shown that this assumption can be improved in future studies by choosing also a non-zero streamwise wavenumber.

\begin{figure}
\centering
\raisebox{1.2in}{(a)}\includegraphics[width=0.45\textwidth]{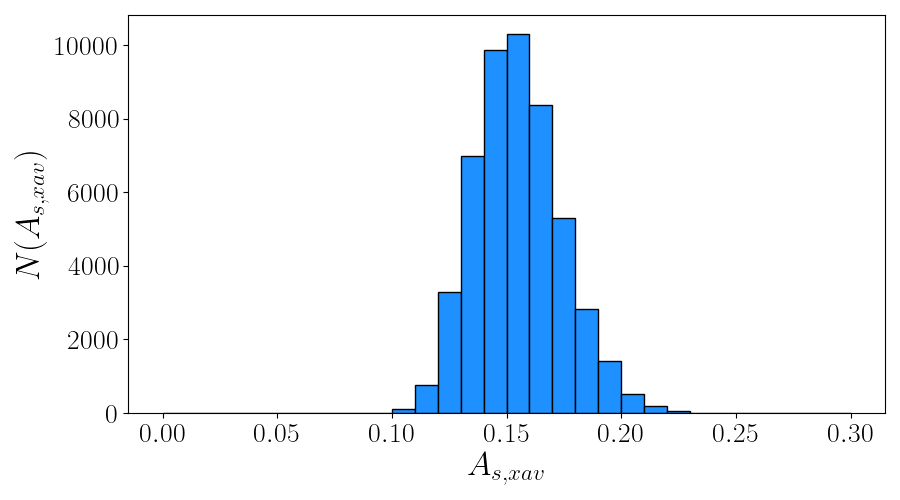}\quad
\raisebox{1.2in}{(b)}\includegraphics[width=0.45\textwidth]{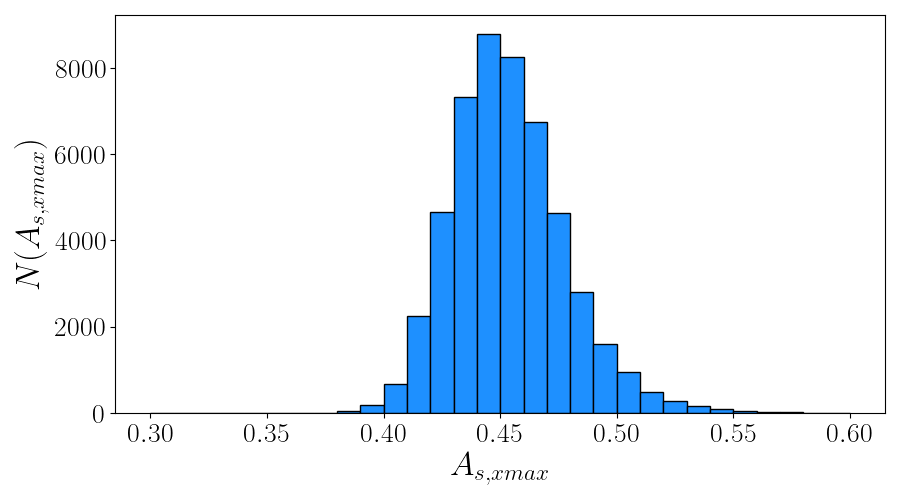}\\
\raisebox{1.2in}{(c)}\includegraphics[width=0.45\textwidth]{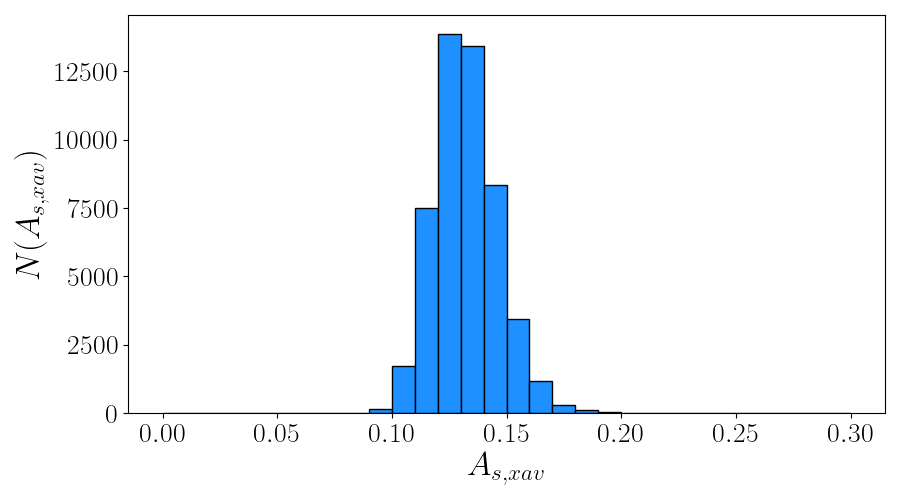}\quad
\raisebox{1.2in}{(d)}\includegraphics[width=0.45\textwidth]{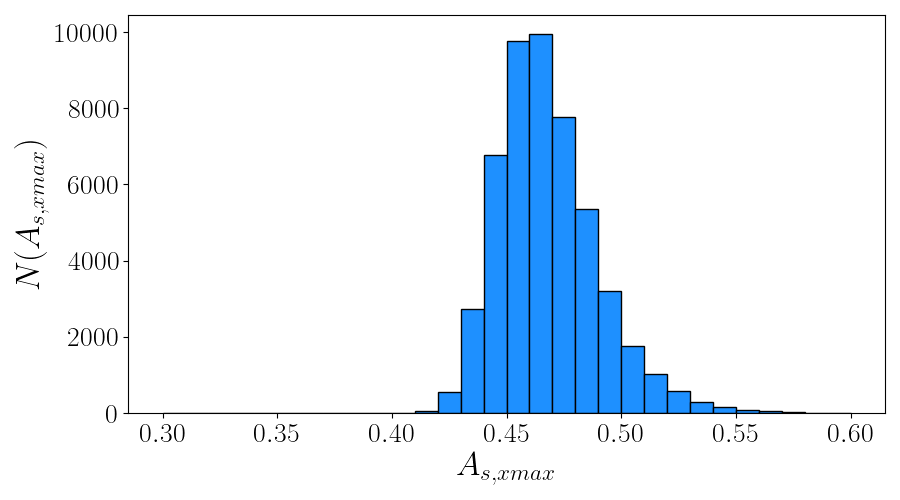}
\caption{Histograms of streak amplitudes measured from DNS in a domain of size $L_x\times L_y\times L_z=35\times2\times15$ for (a-b) $Re_{\tau}=71$ and (c-d) $Re_{\tau}=106$. (a-c) Amplitudes from streamwise averaged fields \eqref{eq:asxav}; (b-d) streamwise maximum amplitudes of non-averaged velocity fields \eqref{eq:asxmax}.}
\label{fig:streaks_amp}
\end{figure}

The values of the critical amplitude reported earlier deserve a discussion in connection with the streak amplitude found either in the literature on streak instability or in DNSs of turbulent flows.
The analysis of the energy budgets in Section \ref{sec:eba} shows that the desired Reynolds dependence is induced by the presence of an eddy viscosity dissipation. 
However, the eddy viscosity dissipation implies also an increase of the critical amplitude of the instability. This fact was already observed by \citet{alizard2015linear}, who compared his critical amplitudes to those of \citet{schoppa2002coherent}, after conversion of his values to the strength factor introduced by the latter. \citet{alizard2015linear} found significantly greater critical strength factors with respect to those from \citet{schoppa2002coherent} and attributed this difference to his choice of eddy viscosity.
Despite the very different $Re$ values, we can directly compare our critical values to those of \citet{alizard2015linear} since we use the same definition of streak amplitude. We find critical amplitudes that are correspondingly larger ($\approx0.40$ {\it versus} $\approx0.18$). Therefore we further investigate this difference using DNS data. DNS in moderate-size periodic domains ($L_x\times L_y\times L_z=35\times2\times15$) have been performed and the following quantities introduced:
\begin{align}
    \label{eq:asxav}
    & A_{s,xav} = \frac{\max_{y,z}\parket{u^{\prime}}_x - \min_{y,z}\parket{u^{\prime}}_x}{2\Ubar_c}, \\[1ex]
    \label{eq:asxmax}
    & A_{s,xmax} = \max_x \frac{\max_{y,z}u^{\prime} - \min_{y,z}u^{\prime}}{2\Ubar_c},
\end{align}
where $u^{\prime}$ is the streamwise velocity fluctuation, $\parket{\cdot}_x$ denotes the streamwise average and $\Ubar_c$ is the centreline velocity of the turbulent mean profile. 
$A_{s,xav}$ measures the amplitude of streamwise-uniform streaks whereas $A_{s,xmax}$ measures the maximum local amplitude of generic streaks. The smoothing effect of the streamwise average implies $A_{s,xav}<A_{s,xmax}$. The distribution of $50,000$ temporal samples of these quantities is shown in figure \ref{fig:streaks_amp} for two $Re$. $A_{s,xav}$ takes values between $0.10$ and $0.30$ whereas $A_{s,xmax}$ takes values between $0.35$ and $0.50$. We conclude that, for the streamwise-uniform streaks of our base flow, a critical amplitude $0.4$ is too high in average. We do not exclude that such an amplitude could be reached locally within a turbulent flow, thereby making a connection with {\it extreme events} \citep{Hack_Schmidt_2021}. The need for large $A_s$ may be a consequence of the various approximations made in the construction of the base flow, notably its over-symmetrisation. It is not excluded that, if the base flow streaks were three-dimensional or characterised by more than one spanwise wavelength, the critical amplitude would decrease, yet this remains to be verified. Appendix \ref{sec:appendix_baseflow_mod} documents the results obtained using a slightly modified base flow. It is shown that the instability is retrieved keeping $A_s$ constant and increasing just the amplitude of the transverse velocity components. This suggests that improving the modelling of the base flow can lower the critical streak amplitude. \\

\begin{figure}
\centering
\includegraphics[width=0.9\textwidth]{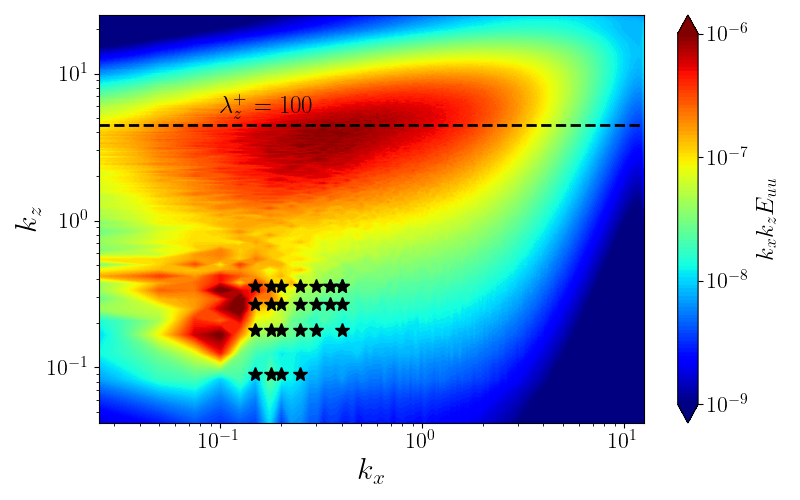}
\caption{Wall-normal integrated pre-multiplied energy spectrum for the streamwise velocity component, time-averaged from DNS at $Re_{\tau} = 71$. The simulation domain is $L_x\times L_y\times L_z = 250\times2\times150$. Black line denotes the structures with spanwise wavelength $\lambda_z^+ = 100$ (as the base flow streaks considered in this work). Black stars denote the large-scale wavelengths of all the unstable modes obtained using the eddy viscosity model, $A_s=0.5$, and having large-scale energy ratio $r_{LS} \geq 2\%$ (see text for detail).}
\label{fig:spectrum_with_modes}
\end{figure}

Notwithstanding the approximations made, the results of the eddy viscosity model support the initial expectation of this work, namely that large-scale structures can emerge out of the instability of a base flow featuring smaller-scale structures. This idea is illustrated in figure \ref{fig:spectrum_with_modes}, which shows the time averaged pre-multiplied energy spectrum of the streamwise velocity component, integrated along the wall-normal direction and obtained from a DNS at $Re_{\tau}=71$ in a large domain with $L_x \times L_z = 250 \times 125$. The flow for $Re$ below critical is characterised by two scales: the streaks and the modulations. There is no marked scale separation between the two, but the two peaks are well discernible in figure \ref{fig:spectrum_with_modes}. The peak corresponding to the streaks is centred around the spanwise wavelength $\lambda_{z,s}^+ = 100$. Streaks also have a characteristic streamwise wavelength yet we have supposed $k_x=0$ as a first approximation. The modulations are characterised by $k_x \approx 0.1$ and $k_z \approx 0.5$, consistently with previous DNS studies \citep{kashyap2020flow,kashyap2022linear}. 
In figure \ref{fig:spectrum_with_modes}, the wavelengths of the unstable modes obtained with the eddy viscosity model at $Re_{\tau}=71$ are superimposed on the DNS spectrum, all unstable modes with $r_{LS}>2\%$ being selected. The range $k_x \le 0.5$ was considered, but unstable modes with $r_{LS}>2\%$ were found only for $k_x\in[0.15,0.4]$. The spanwise wavelength of the eigenmode was extracted using spanwise Fourier transform. The unstable modes are found in a region of the spectrum close to where the large-scale turbulent modulations lie. Closer inspection reveals that they cover very well the spanwise wavenumber range of the modulations but less accurately the streamwise wavenumber range. Note that by assumption, the base flow is invariant in $x$, therefore the unstable mode is monochromatic in $x$. As discussed above, this is a first approximation. The results may be improved provided characteristic  streamwise wavelengths of the streaks are taken into account in the base flow. This improvement is not straightforward from the computational point of view but remains a stimulating perspective for future studies.\\

The present modelling effort is directed specifically at the analysis of the linear instability of the turbulent flow, in line with the studies of \citet{kashyap2022linear,kashyap2024linear}. As such, the model is representative of the physics only in the incipient modulational regime, when the solution does not depart sensibly from the turbulent attractor. However the link between the oblique structure shown in figure  \ref{fig:eigenmode_xz} and genuine laminar-turbulent patterns, although visually stimulating, is not trivial. Both the eigenmode and the base flow are three-dimensional vector fields and the result of adding them together is not simple. In particular, active parts of the eigenmode can either tame the turbulence or reinvigorate it.

Fully developed laminar-turbulent patterns remain thus outside the scope of the present model. Indeed, laminar flow is not a solution of the nonlinear model equations \eqref{eq:sns}.
Therefore, nonlinear simulations of the instability following Eq. \eqref{eq:sns} are not expected to be physically relevant to laminar-turbulent patterning, given that the flow would not have the possibility to develop the laminar holes observed in DNS \citep{kashyap2020flow}. Nonlinear simulations of \eqref{eq:sns} have been performed anyway and they confirm this expectation : the oblique stripe saturates in amplitude but no proper laminar hole can develop because the streaks are constantly forced everywhere in the domain (not shown). An improved model capturing the linear instability while allowing for both laminar and turbulent solutions is the object of current investigation. The recent efforts of \citet{kashyap2025laminar} and \citet{benavides2025model} are a useful starting point in this sense.   


\section{Conclusions}

This work is motivated by the recent observation that laminar-turbulent patterns in channel flow emerge, as the Reynolds number is decreased, from spatial modulations that themselves arise as a linear instability of the underlying turbulent flow \citep{kashyap2022linear}. This {\it top-to-bottom} point of view with decreasing $Re$ differs from the {\it bottom-up} description with increasing $Re$ \citep{duguet2024puffing}, although there is no incompatibility between the two. Our work reveals that the modulational instability modulations observed in turbulent channel flow can be modeled as a detuned instability of a mean flow superimposed with streamwise streaks.

For consistency with the observations from DNS, the effect of the turbulent background needs to be taken into account through an eddy viscosity closure. The energy budget analysis  shows that a critical Reynolds number is predicted by the model owing to the Reynolds dependence of the eddy viscosity. This  suggests that the pattern-forming instability is inherent to the turbulent flow, in accordance with the ideas in \citet{kashyap2022linear}, the difference being that some knowledge of the coherent structures of the turbulent flow (the streaky part) is also incorporated. 
Modelling linear instabilities in turbulent environments is never trivial and several assumptions need to be made, including the choice for a closure. We have discussed some of the most delicate points in our present formulation and have suggested directions for future improvements. Beyond the need for better modeling, future studies should address the question of generalisation to other wall-parallel flows like plane Couette flow, Taylor-Couette flow, Couette-Poiseuille flow and the Asymptotic Suction Boundary Layer (ASBL). At present, it is not clear whether the linear modulational instability discovered for the channel flow is equally relevant for these other configurations. The subject of pattern formation in turbulent shear flows remains full of opportunities for future studies.\\

\backsection[Acknowledgements]{The authors would like to thank C. Cossu for useful discussions. N.C. was funded by the Italian Ministry of University and Research. Y.D. thanks the Dynfluid laboratory for its hospitality, while N.C. thanks A. Franchini for useful discussions. This work was granted access to the HPC resources of TGCC under the allocation 2024-[A0152A06362] and 2025-[A0172A06362] made by GENCI.}

\backsection[Declaration of interests]{The authors report no conflict of interest.}


\begin{figure}
\centering
\raisebox{2.8in}{(a)}\includegraphics[width=0.45\textwidth]{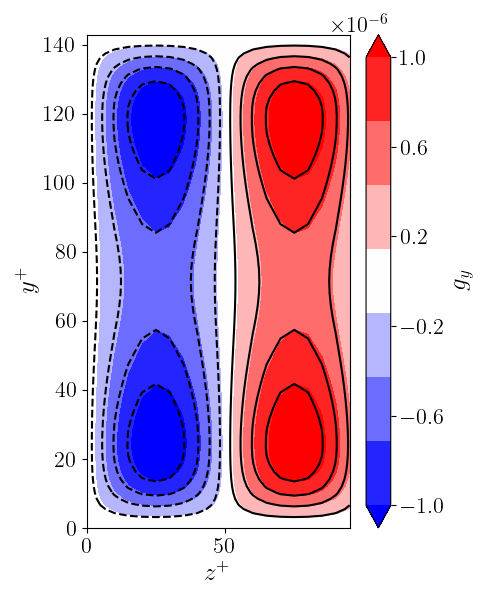}
\raisebox{2.8in}{(b)}\includegraphics[width=0.45\textwidth]{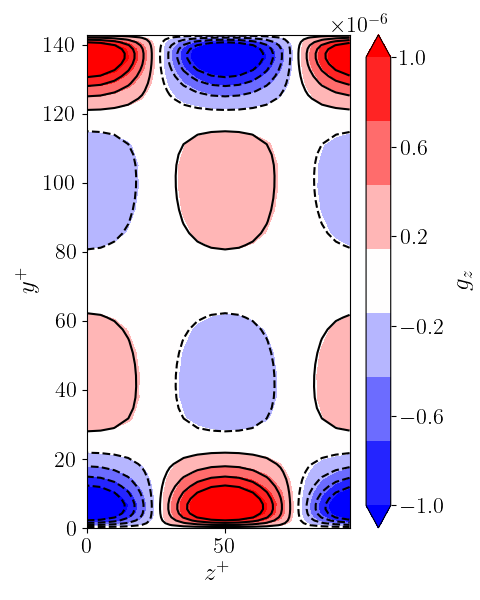} \\
\raisebox{2.2in}{(c)}\includegraphics[width=0.8\textwidth]{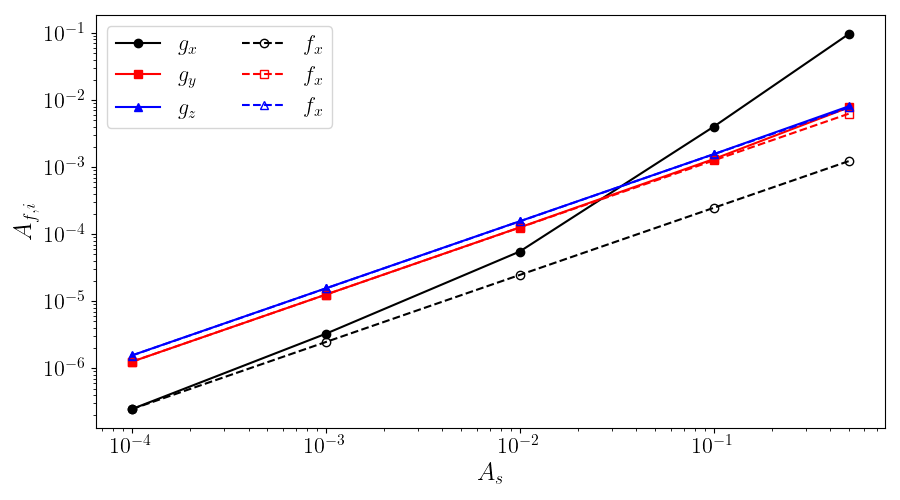}
\caption{(a-b): Forcing term $\vec{g}$ in equations (\ref{eq:fixed-point}-\ref{eq:sns}) (shaded contours) compared to the forcing term $\vec{f}$ in equation \eqref{eq:eddy-resolvent} (black lines: solid for positive values, dashed for negative values) for $A_s=1\times10^{-4}$. (a) Wall-normal component; (b) spanwise component. The line contours have the same iso-levels of the shaded contours to make the comparison meaningful. (c): Comparison of the streak forcing amplitude ($A_{f,i}=(\max_{y,z} f_i - \min_{y,z} f_i)/2$) as a function of the streak amplitude ($A_s$). $\vec{f}$ and $\vec{g}$ defined as in panels (a-b).}
\label{fig:forcing}
\end{figure}

\appendix 

\section{Computation and validation of the streak forcing term}
\label{sec:appendix_forcing}

The forcing $\vec{g}$ in the system \eqref{eq:sns} is numerically computed using the nonlinear time-stepping code \textsf{channelflow} \citep{gibson2021}. To detail the procedure, let us denote the left-hand side of the first equation in \eqref{eq:sns} as $\partial \vec{u}/\partial t + \vec{\mathcal{N}}(\vec{u})$. We seek $\vec{g}$ such that, by construction, $\vec{U}$ verifies \eqref{eq:fixed-point}, {\it i.e.} $\vec{\mathcal{N}}(\vec{U})=\vec{g}$. To approximate $\vec{\mathcal{N}}(\vec{U})$ using a timestepper, and only for this purpose, we consider the initial value problem:
\begin{equation}
    \begin{cases}
        \displaystyle \derpar{\vec{u}}{t} = - \vec{\mathcal{N}}(\vec{u}), \\[1ex]
        \displaystyle \diver{\vec{u}}=0, \\[1ex]
        \displaystyle \vec{u}(t=0) = \vec{U}.
    \end{cases}
\end{equation}
Advancing this system of one time step $\Delta t$, one obtains a velocity field $\vec{U}^1 \neq \vec{U}$. Then, the forcing term $\vec{g}=\vec{\mathcal{N}}(\vec{U})$ can be approximated by the finite difference $-\parton{\vec{U}^1-\vec{U}}/\Delta t$.

This approach comes with a numerical error due to the finite difference approximation. In practice, however, it turns out accurate enough. Indeed, advancing the system \eqref{eq:sns} with the timestepping code starting from $\vec{U}$, this field remains steady up to a small numerical error, as expected by construction of the forcing. Moreover, simulations of the linear instability of the streaky base flow have been performed using the nonlinear code and the growth rate predicted by the stability analysis was retrieved up to a $<5\%$ error (not shown). Therefore, the linear and nonlinear codes are consistent and the numerical procedures are accurate enough for the purposes of this work.

Another way to validate the forcing computed with the timestepper code is to consider small $A_s$. In the limit $A_s \rightarrow 0$, equations \eqref{eq:eddy-resolvent} and \eqref{eq:fixed-point} coincide with $\vec{u}^{\prime}=\vec{u}_s$ and $\vec{g} \rightarrow \vec{f}$ because the difference between the equations $A_s^2\vec{u}_s\cdot\grad{\vec{u}_s}$ becomes negligible. This is shown in panels (a) and (b) of figure \ref{fig:forcing} for $A_s=1\times10^{-4}$. As $A_s$ increases, the two forcings differ, especially on the streamwise component (figure \ref{fig:forcing} (c)). This difference at finite $A_s$ motivates the need for a specific procedure to compute $\vec{g}$.

\begin{figure}
\centering
\raisebox{1.4in}{(a)}\includegraphics[width=0.45\textwidth]{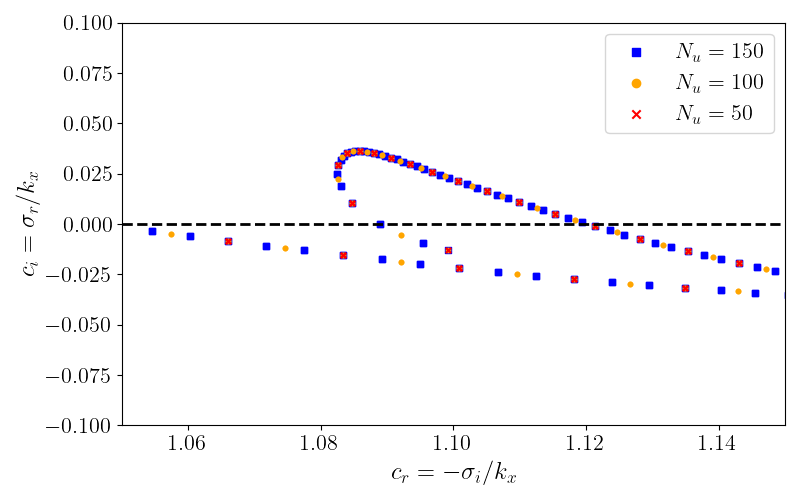}\quad
\raisebox{1.4in}{(b)}\includegraphics[width=0.45\textwidth]{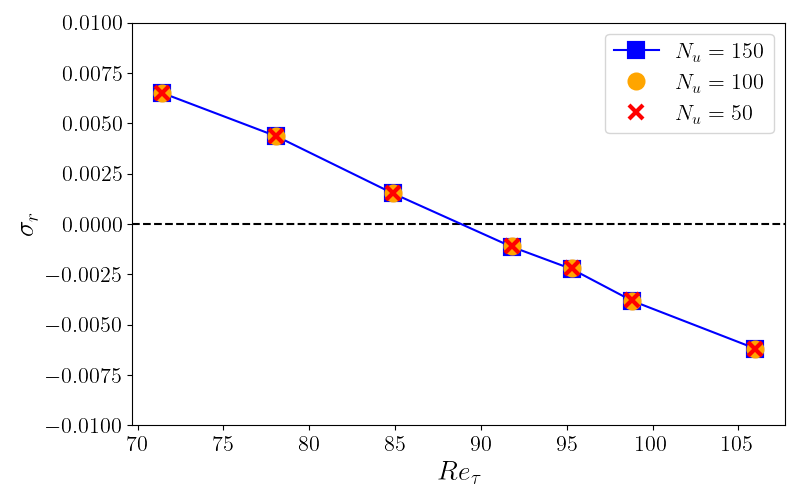}
\caption{Convergence of the eigenvalues with the number $N_u$ of base flow units (eddy viscosity model). (a) Eigenvalues for $Re_{\tau}=71$, $k_x=0.18$ and $A_s=0.5$. (b) Leading growth rate as a function of Reynolds number for $k_x=0.18$ and $A_s=0.5$.}
\label{fig:check_Nu}
\end{figure}

\begin{figure}
\centering
\raisebox{1.4in}{(a)}\includegraphics[width=0.45\textwidth]{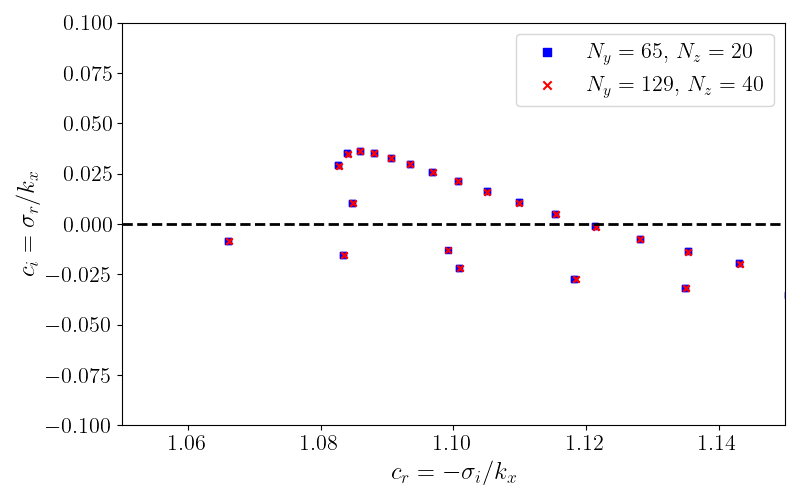}\quad
\raisebox{1.4in}{(b)}\includegraphics[width=0.45\textwidth]{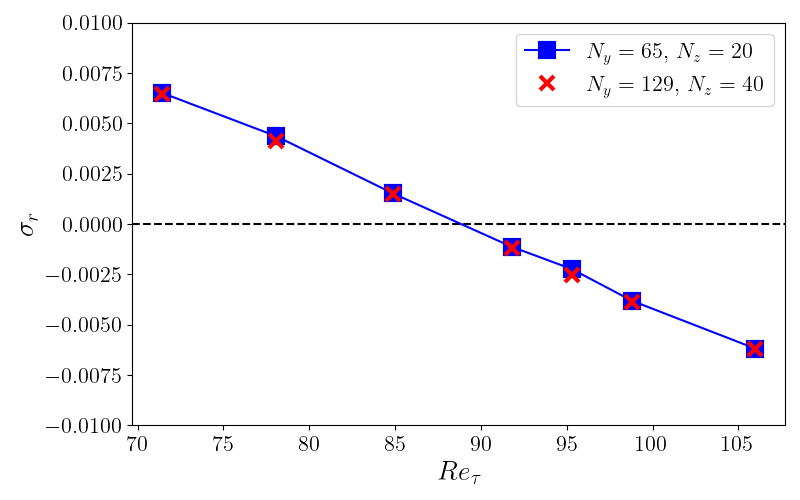}
\caption{Convergence of the eigenvalues with the number of collocation points (eddy viscosity model). (a) Eigenvalues for $Re_{\tau}=71$, $k_x=0.18$ and $A_s=0.5$. (b) Leading growth rate as a function of $Re$ for $k_x=0.18$ and $A_s=0.5$.}
\label{fig:check_grid}
\end{figure}

\section{Convergence of the eigenvalues}
\label{sec:appendix_convergence}

In the main text, the dependence of the stability results on key parameters like the streak amplitude $A_s$, the Reynolds number $Re_{\tau}$ and the streamwise wavenumber $k_x$ was thoroughly analysed. The stability analysis, however, involves also numerical parameters like the number of collocation points for the discretisation of the problem ($N_y$ in the wall-normal direction and $N_z$ in the spanwise direction) and the number of repeated base flow units which effectively imposes the spanwise size of the domain ($N_u$). This appendix shows that, as expected, the results are independent of these parameters when their values are sufficiently high.

Figure \ref{fig:check_Nu}(a) shows the effect of increasing the number of independent base flow units on the spectrum: the branches of detuned modes tend towards continuous branches. However, as demonstrated by figure \ref{fig:check_Nu}(b), the leading eigenvalue (and corresponding mode) is already well captured with $N_u=50$, which means that the branch is sufficiently well discretised for this number of units.

In a similar way, figure \ref{fig:check_grid}(a) shows that doubling the number of collocation points in both spatial directions affects minimally the eigenvalues. The same is true for the spatial structure of the eigenmodes (not shown). Moreover, figure \ref{fig:check_grid}(b) shows that the $Re$-dependence of the leading growth rate is not affected either.

Therefore, the choice of $N_y=65$, $N_z=20$ and $N_u=50$ employed for all the computations analysed in the article does not affect the conclusions of this work.

\section{Three-dimensional streaks}
\label{sec:appendix_3Dstreaks}

\begin{figure}
\centering
\raisebox{1.5in}{(a)}\includegraphics[width=0.46\textwidth]{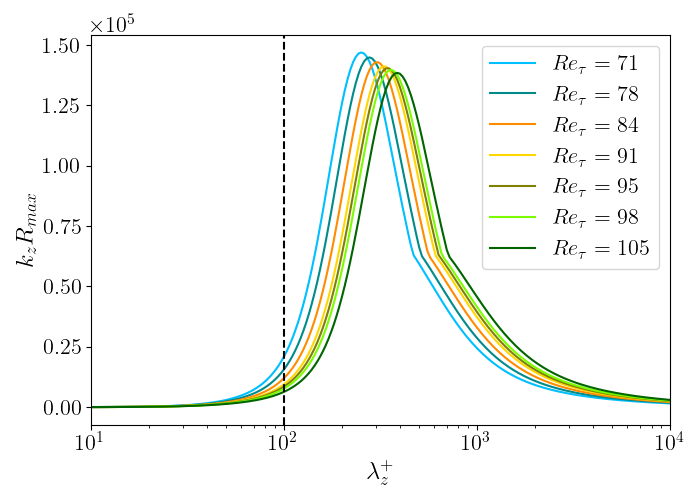}\quad
\raisebox{1.5in}{(b)}\includegraphics[width=0.44\textwidth]{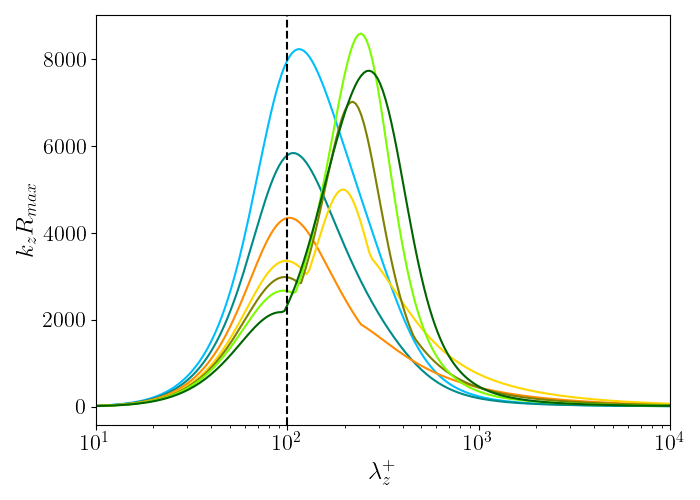} \\
\caption{Maximum energy amplification factor as a function of the spanwise wavelength of the forcing (in wall units, $\lambda^+_z$) for streamwise wavenumber (a) $k_x=0$ and (b) $k_x^+=2\pi/1000$. }
\label{fig:resolvent_gain}
\end{figure}

In the main text, streaks are computed maximizing the resolvent norm for $k_x=0$ and $\lambda_{z,s}^+=100$. However, if the maximum amplification factor (cf. \eqref{eq:resnorm})
\begin{gather}
    R_{max}(k_x,k_z) = \max_{\omega} R(k_x,k_z,\omega),
\end{gather}
is considered as a function of $\lambda_z=2\pi/k_z$ for $k_x=0$, then the maximum amplification does not correspond to $\lambda_z^+=100$ (see figure \ref{fig:resolvent_gain} (a)). 
Indeed, the curve is characterised by only one peak whose wavelength scales in outer rather than inner units. In contrast, at higher $Re_{\tau}$, there are two peaks, one of which scales in wall units and corresponds to $\lambda_z^+ \approx 100$ (see the results of \citet{hwang2010linear}, reproduced successfully, not shown here). 
As $Re_{\tau}$ is decreased the peak in wall units disappears and the curve displays only one peak which does not match $100$ wall units. This behaviour was also documented in other studies of optimal harmonic forcing at relatively low $Re$ \citep{hwang2010b,willis2010,pujals2010}.
One possible explanation is the strong hypothesis $k_x=0$. Streaks are elongated in the streamwise direction, hence $k_x=0$ is a decent first approximation.
Streaks in turbulent flow, however, are characterised by a distribution of (finite) streamwise wavelengths which, according to several observations, is centered around $\lambda_x^+\approx 1000$ wall units \citep{smits2011}.
Figure \ref{fig:resolvent_gain} (b) shows that, when the optimal amplification factor curves are shown for $k_x^+=2\pi/1000$, the peak at $\lambda_z^+=100$ is recovered.
For the purpose of this work, this implies that the choice of a finite streamwise wavelength for the streaks would be more relevant.
Moreover, the value of $100$ inner units represents in real flows only the mean spanwise wavelength \citep{kline1967structure}. An alternative would be the generalisation to a three-dimensional base flow featuring three-dimensional streaks. Such a choice is technically possible, however, it would make the stability computations more expensive and would require different numerical methods. It is therefore outside the scope of the present work.

\section{Instability of a modified base flow}
\label{sec:appendix_baseflow_mod}

\begin{figure}
\centering
\raisebox{1.4in}{(a)}\includegraphics[width=0.45\textwidth]{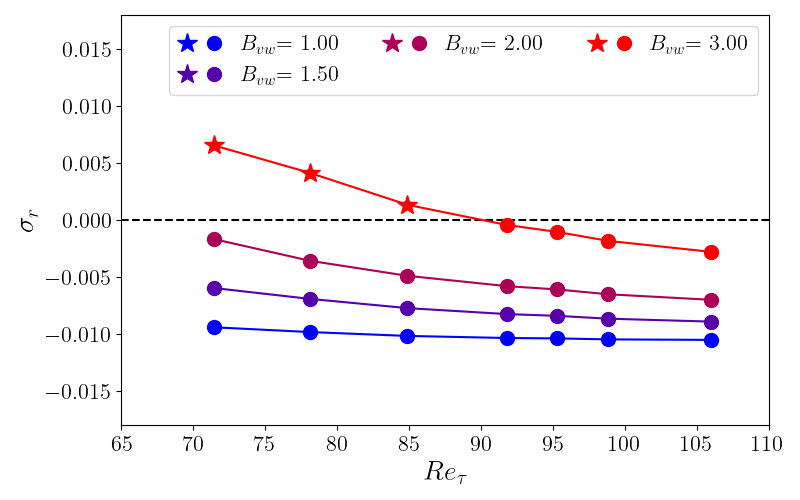}\quad
\raisebox{1.4in}{(b)}\includegraphics[width=0.45\textwidth]{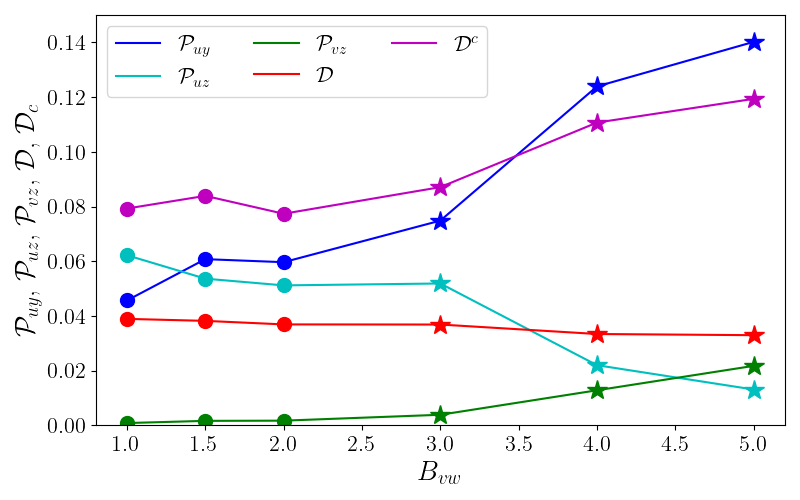}
\caption{Effect of increasing the transverse components amplitude in the base flow. (a): Leading growth rate as a function of $Re$ for $k_x=0.18$. $A_s$ is kept constant at $0.2$ but the transverse velocity components of the base flow $V$ and $W$ are multiplied by a factor $B_{vw}$. (b): Variation of the energy budgets with the prefactor $B_{vw}$ ($A_s=0.2$, $Re_{\tau}=71$, $k_x=0.18$). For the meaning of the labels see equations (\ref{eq:globalOR}-\ref{eq:Cterm}).}
\label{fig:vwboost}
\end{figure}

The leading resolvent modes have energy predominantly in the streamwise velocity component and almost no energy in the transverse components \citep{hwang2010linear}. However, in actual turbulent flows, the velocity field possesses three genuine components, at least because in the SSP picture streamwise streaks are accompanied by streamwise vortices. In order to analyse the influence of this three-dimensionality on critical thresholds, a numerical experiment was performed (using the eddy viscosity model) by augmenting the amplitude of transverse components of the vector field $\vec{u}_s$ while $A_s$ was kept constant. This is easily implemented since the streamwise invariance of the field $\vec{u}_s$ implies that streamwise and transverse components are decoupled in the continuity equation. Therefore if, after the rescaling of $\vec{u}_s$, $v_s$ and $w_s$ are multiplied by a factor $B_{vw}$, the resulting velocity field is still divergence-free. The dependence of the leading growth rate on $Re_{\tau}$ is shown in figure \ref{fig:vwboost}(a) for different values $B_{vw}\geq1$. When $B_{vw}$ is large enough, a critical Reynolds number appears exactly as in figure \ref{fig:gr_vs_Re}(b). Thus, with a slightly modified base flow, the right $Re$-dependence of the growth rates can be retrieved also for $A_s=0.2$, which is comparable with other streak instabilities critical amplitudes \citep{park2011,alizard2015linear,ciola2024large}. Importantly, the augmented amplitudes of the transverse components are of the order of the mean amplitudes of the transverse fluctuations observed in DNS (not shown). Therefore, the modified base flow does not contain any extreme fluctuations in contrast to the previous base flow with $A_s=0.5$.  
The energy budgets obtained using the new base flow (figure \ref{fig:vwboost} (b)) show similar dissipation trends with respect to those in figure \ref{fig:eb_vars}(b) but different trends on the production contributions, with the production term $\mathcal{P}_{vz}$ playing a non-negligible role for large $B_{vw}$. This numerical experiment shows that certain quantitative aspects of the stability analysis depend on explicit choices in the modelling of the base flow. By contrast, the results highlighted in the main text are qualitatively robust to minor modifications of the base flow.

\bibliographystyle{jfm}
\bibliography{bibliography}

\end{document}